\documentclass[12pt]{article}
\pdfoutput=1

\usepackage{graphicx,psfrag,epsf,color}
\usepackage{amsmath,amssymb,amsfonts}
\usepackage{array}
\usepackage{cite}
\usepackage{rotating}
\usepackage{slashed, cancel}
\usepackage{scalerel,stackengine}

\numberwithin{equation}{section}

\newcounter{MBQ}

\def\slash#1{#1 \hskip-0.45em /}

\newcommand{\be}{\begin{equation}}
\newcommand{\ee}{\end{equation}}
\newcommand{\bea}{\begin{eqnarray}}
\newcommand{\eea}{\end{eqnarray}}
\newcommand{\bi}{\begin{itemize}}
\newcommand{\ei}{\end{itemize}}
\newcommand{\ben}{\begin{enumerate}}
\newcommand{\een}{\end{enumerate}}
\newcommand{\bt}{\begin{tabular}}
\newcommand{\et}{\end{tabular}}

\newcommand{\nn}{\nonumber}

\newcommand{\nm}{n_-}
\newcommand{\np}{n_+}

\global\long\def\order#1{\mathcal{O}\left(#1\right)}

\newcommand\sA{\ThisStyle{\ensurestackMath{%
 {%
\stackinset{r}{}{c}{}{\SavedStyle/}{\SavedStyle\mathcal{A}}}}}}

\definecolor{darkgreen}{rgb}{0.0,0.6,0.0}

\begin{document}
\allowdisplaybreaks

\begin{titlepage}

\begin{flushright}
{\small
TUM-HEP-1164/18\\
NIKHEF/2018-046\\
arXiv:1809.10631 [hep-ph]\\
February 08, 2019
}
\end{flushright}

\vskip1cm
\begin{center}
{\Large \bf\boldmath Leading-logarithmic threshold resummation 
of the \\[0.2cm]  
Drell-Yan process
at next-to-leading power}
\end{center}

\vspace{0.5cm}
\begin{center}
{\sc Martin~Beneke},$^{a}$ 
{\sc Alessandro Broggio},$^{a}$
{\sc Mathias~Garny},$^{a}$\\ 
{\sc Sebastian Jaskiewicz},$^{a}$ 
{\sc Robert~Szafron},$^{a}$,
{\sc Leonardo Vernazza},$^{b,c}$\\ 
and {\sc Jian~Wang}$^{a}$\\[6mm]
{\it $^a$ Physik Department T31,\\
James-Franck-Stra\ss e~1, 
Technische Universit\"at M\"unchen,\\
D--85748 Garching, Germany\\[0.2cm]
}
{\it $^b$ Institute for Theoretical Physics Amsterdam \\
and Delta Institute for Theoretical Physics, \\
University of Amsterdam, Science Park 904,\\
NL--1098 XH Amsterdam, The Netherlands\\[0.2cm]
}
{\it $^c$ Nikhef, Science Park 105,\\ 
NL--1098 XG Amsterdam, 
The Netherlands\\
}
\end{center}

\vspace{0.6cm}
\begin{abstract}
\vskip0.2cm\noindent
We resum the leading logarithms $\alpha_s^n \ln^{2 n-1}(1-z)$, 
$n=1,2,\ldots$ near the kinematic threshold $z=Q^2/\hat{s}\to 1$ of 
the Drell-Yan process at next-to-leading power in the expansion 
in $(1-z)$. The derivation of this result employs 
soft-collinear effective theory in position space and 
the anomalous dimensions of subleading-power soft functions, 
which are computed. Expansion of the resummed result leads to the 
leading logarithms at fixed loop order, in agreement with 
exact results at NLO and NNLO and predictions from the physical 
evolution kernel at N$^3$LO and N$^4$LO, and to new results 
at the five-loop order and beyond. 
\end{abstract}
\end{titlepage}

\section{Introduction}
\label{sec:introduction}

The weak-coupling expansion of QCD high-energy scattering fails near 
kinematic thresholds due to the restricted phase space for real 
emission. The logarithmic enhancements in the kinematic variable 
that characterizes the threshold must be resummed to all orders 
in the coupling expansion to arrive at a reliable approximation. 
This has been studied first \cite{Sterman:1986aj,Catani:1989ne} and 
in greatest detail 
for the simplest such situation, the production of a single uncoloured 
particle DY (Drell-Yan process) in the collision of two hadrons, 
$A(p_A) B(p_B) \to \mbox{DY}(Q)+X$, where $X$ denotes an unobserved 
QCD final state. The DY process has always provided the first physically 
very relevant case on which to push the accuracy of resummation to 
the next level, or explore new approaches to 
resummation \cite{Becher:2007ty}.

The DY spectrum $d\sigma_{\rm DY}/dQ^2$ is given by the convolution of 
parton distributions in the incoming hadrons with partonic short-distance 
cross sections $\hat{\sigma}_{ab}$ in partonic channels $ab$. The parton 
scattering cross sections can be regarded as functions of $z=Q^2/\hat{s}$,  
where $\hat{s} = x_a x_b s$ is the partonic center-of-mass (cms) energy 
squared, and 
$x_a,x_b$ the momentum fractions of the partons in the corresponding 
hadrons. Near the partonic threshold $z=1$, $\hat{\sigma}_{ab}$ has the 
singular expansion
\begin{equation}
\hat{\sigma}_{ab}(z) = \sum_{n=0}^\infty 
\alpha_s^n\left[ c_n\delta(1-z)+\sum_{m=0}^{2 n-1}\left(
c_{nm}\left[\frac{\ln^m(1-z)}{1-z}\right]_+ 
+d_{nm} \ln^m(1-z)\right) +\ldots\right].
\end{equation}
In this expression the series with coefficients $c_n, c_{nm}$
encompass the {\it leading power} (LP) singular terms, and, more 
specifically, the terms $c_0$ and $c_{n (2n-1)}$ constitute the 
{\em leading logarithms} (LL). The terms multiplied by $d_{nm}$ 
are suppressed by one power of $(1-z)$ and are referred to as 
{\it next-to-leading power} (NLP). The NLP LL series is given by 
the highest power NLP logarithms with coefficients $d_{n (2n-1)}$ 
for $n=1,2,\ldots$.

Existing approaches to soft gluon resummation of the DY threshold 
apply only to the LP terms. The key result is the factorization of 
the partonic cross section
\begin{equation}
\hat{\sigma}(z) = H(Q^2) \,Q S_{\rm DY}(Q(1-z))
\label{eq:LPfact}
\end{equation}
into the product of a hard function and the DY soft function 
\cite{Korchemsky:1993uz}
\begin{equation}
S_{\rm DY}(\Omega) = \int \frac{dx^0}{4\pi}\,e^{i x^0 \Omega/2}\,
\frac{1}{N_c}\,\mbox{Tr} \,
\langle 0|\mathbf{\bar{T}}(Y^\dagger_+(x^0) Y_-(x^0)) 
\,\textbf{T}(Y^\dagger_-(0) Y_+(0))
|0\rangle
\label{eq:LPsoftfn}
\end{equation}
expressed in terms of Wilson lines, as defined below. Both functions 
depend on a renormalization scale $\mu$. This 
dependence is important to perform the resummation via a renormalization 
group equation, but will not be indicated explicitly unless 
necessary. In principle it is possible to sum arbitrary subleading 
logarithms at LP by computing the hard and soft function and the 
evolution equation to sufficiently high order. Presently, LP 
logarithms can be summed to the next-to-next-to-next-to-leading 
logarithmic order \cite{Moch:2005ky,Becher:2007ty}.

In contrast, much less is understood at NLP. The structure of NLP 
logarithms has recently received increased interest with explicit 
calculations at fixed order $n=1,2$ using the method of region 
approach \cite{Bonocore:2014wua,Anastasiou:2014lda,Bahjat-Abbas:2018hpv} 
and diagrammatic factorization techniques 
\cite{Bonocore:2015esa,Bonocore:2016awd,DelDuca:2017twk}. 
However, an all-order resummation has not yet been performed for 
NLP threshold logarithms in the Drell-Yan process.

In the present work we accomplish this task for the leading logarithmic 
terms at NLP for the quark-antiquark production channel of 
the DY process within the framework of soft-collinear 
effective theory (SCET). This framework has the distinct advantage 
of providing precise operator definitions of the factors appearing 
in the factorization formula extended to NLP, which converts the 
resummation problem into a renormalization problem of SCET operators 
and soft functions. In Sec.~\ref{sec:NLPfact} we discuss the 
factorization formula for the DY process at NLP. While the form of 
the result is rather intuitive, this section draws heavily on a 
companion publication \cite{BBJVunpublished}, where this formula 
is derived in detail and verified at the one-loop order. The core result 
of NLP LL resummation is contained in Sec.~\ref{sec:LLresum}, 
which identifies the sources of NLP LLs and derives the 
hard, soft and collinear functions needed for resummation, as 
well as the renormalization-group equation and its LL solution. 
In Sec.~\ref{sec:FOexpansion} we expand the resummed result in 
$\alpha_s$, which provides both, a check of the resummed result by comparison 
with existing fixed-order expressions, and the so far unknown logarithmic 
terms at higher order. 
The technical details of the one-loop anomalous dimension calculation 
for the soft functions and the derivation of an evolution 
equation for the NLP partonic cross section are given in two appendices.

\section{Factorization near threshold at NLP}
\label{sec:NLPfact}

The following treatment makes use of SCET \cite{Bauer:2000yr,Bauer:2001yt} 
in the position-space representation \cite{Beneke:2002ph,Beneke:2002ni}. 
In order to describe the Drell-Yan threshold SCET must include 
collinear, anticollinear and soft fields with momentum 
scaling $Q(1,\lambda^2,\lambda)$,  $Q(\lambda^2,1,\lambda)$
and $Q (\lambda^2,\lambda^2,\lambda^2)$, respectively, 
where the components refer to 
$(\np k,\nm k,k_\perp)$ with $\np^2=\nm^2=0$, $\np\nm=2$, and the 
power counting parameter is $\lambda=\sqrt{1-z}$.  
In addition to the above (anti)collinear modes with parametric 
threshold scaling, the ordinary parton distributions are defined in terms of 
(anti)collinear-PDF modes with scaling $(Q,\Lambda^2/Q,\Lambda)$ 
for c-PDF, and  $(\Lambda^2/Q,Q,\Lambda)$ for 
$\bar{c}$-PDF,\footnote{The above classification of regions refers to the 
`standard' treatment of factorization near threshold, which neglects 
momentum regions which lead to scaleless integrals in dimensional 
regularization. If one aims at the correct identification of ultraviolet 
and infrared singularities, the situation is more subtle, as discussed 
in \cite{Chay:2017bmy}. The whole effect of the additional collinear-soft 
region introduced in this reference is to convert the IR singularities in 
the soft function of the standard treatment into UV singularities, 
ultimately leading to the same factorization formula at leading power. 
If this holds at LP, it must also hold at NLP, since the additional 
region interpolates between the soft scale and the scale of the 
parton distributions, which are the same leading-twist parton 
distributions at NLP in the threshold expansion.}
where $\Lambda$ denotes the strong interaction scale.

In the following we summarize some general results on the 
DY threshold at NLP from~\cite{BBJVunpublished}. We first  
note from \eqref{eq:LPfact} that the LP factorization formula does 
not contain a collinear function with threshold momentum scaling for two 
reasons:  1) Due to threshold kinematics generic collinear modes 
cannot be radiated into the hadronic final state. 2) Virtual collinear 
loops are scaleless, because a threshold-collinear scale can only be formed by 
attaching a soft momentum to the collinear loop. However, after the 
soft-gluon decoupling transformation \cite{Bauer:2001yt} of the collinear 
fields, there are no soft-collinear interactions in the leading-power 
SCET Lagrangian. 

The factorization of the DY threshold at NLP is obtained by matching the 
coupling to the DY particle to higher order in the $(1-z)$ expansion and by 
including subleading interactions from the SCET Lagrangian as 
perturbations. The decoupling of collinear and soft fields in the 
LP Lagrangian guarantees that the amplitude factorizes into a 
convolution of hard, collinear and soft amplitudes. However, 
reason 2) no longer holds at NLP, since the NLP SCET Lagrangian 
contains soft-collinear interactions. The time-ordered products 
with the hard vertex inject soft momentum into the collinear loops, 
resulting in a non-vanishing collinear function {\em at the amplitude 
level}. 

The DY spectrum for the production of a lepton-antilepton pair with 
invariant mass $Q$ through a virtual photon can be written as 
(leaving out quark electric charge factors)
\begin{equation}
\frac{d\sigma_{\rm DY}}{dQ^2} = 
\frac{4\pi\alpha_{\rm em}^2}{3 N_c Q^4}
\sum_{a,b} \int_0^1 dx_a dx_b\,f_{a/A}(x_a)f_{b/B}(x_b)\,
\hat{\sigma}_{ab}(z)\,.
\label{eq:dsigsq2}
\end{equation}
The factorization formula for $\hat{\sigma}_{ab}(z)$ for $z\to 1$ covering 
power-suppressed terms in $(1-z)$ is
\begin{eqnarray}
\hat{\sigma}(z) &=&
\sum_{\rm terms} 
\int d\omega_i d\bar{\omega}_i d\omega_i^\prime d\bar{\omega}_i^\prime\,
D(-\hat{s};\omega_i,\bar{\omega}_i)
D^*(-\hat{s};\omega_i^\prime,\bar{\omega}_i^\prime)
\nonumber\\
&&\times \,Q^2 \int\frac{d^3\vec{q}}{(2\pi)^3 \,2\sqrt{Q^2+\vec{q}^{\,2}}}\,
\frac{1}{2\pi} \int d^4x\,
e^{i(x_a p_A+x_b p_B-q)\cdot x}\, 
\widetilde{S}(x;\omega_i,\bar{\omega}_i,\omega_i^\prime,
\bar{\omega}_i^\prime)\,.\qquad
\label{eq:NLPfact}
\end{eqnarray}
The formula holds for each partonic channel $ab$, and we dropped 
the parton indices $ab$ for notational convenience. We also used the 
same symbol $i$ for four separate index {\it sets} $\,i_j$ and a sum over 
various terms of this form is implicit. This will be made more 
precise in the following section for the terms relevant to LL 
resummation. $\widetilde{S}(x;\omega_i,\bar{\omega}_i,\omega_i^\prime,
\bar{\omega}_i^\prime)$ denotes a generalized multilocal soft function, 
which depends on the soft momenta radiated from the collinear 
and anticollinear functions. The coefficient function 
$D(-\hat{s};\omega_i,\bar{\omega}_i)$ is defined at the amplitude level, 
and summarizes convolutions of the hard functions with the initial-state 
collinear and anticollinear functions at the amplitude level:
\begin{eqnarray}
D(-\hat{s};\omega_i,\bar{\omega}_i) &=& 
\int d(\np p_i) d(\nm \bar{p}_i) \,C(\np p_i,\nm \bar{p}_i)\,
\nonumber\\
&&\times\,J(\np p_i,x_a \np p_A;\omega_i)\,
\bar{J}(\nm \bar{p}_i,-x_b \nm p_B;\bar{\omega}_i)\,.
\label{eq:Dfn}
\end{eqnarray}
Here $C(\np p_i,\nm \bar{p}_i)$ is the momentum-space coefficient function of 
a general two-jet operator as defined in \cite{Beneke:2017ztn}. 
Beyond LP, its Fourier transform can depend on several momentum fractions 
as indicated by the unspecified generic index $i$. 
Note that colour and Lorentz indices on the hard, collinear and 
soft functions are suppressed for the purpose of this generic discussion.

The LP factorization formula is recovered from the general formula 
as the special case when there are no collinear functions. In this 
case the index set $i$ is empty, the convolutions over the various 
$\omega_i$ variables in \eqref{eq:NLPfact} are absent and 
$D(-\hat{s};\omega_i,\bar{\omega}_i)\to C^{A0}(-\hat{s}) \equiv 
C^{A0}(x_a\np p_A,x_b \nm p_B)$, where the 
latter denotes the matching coefficient of the LP SCET current, 
\begin{equation}
\bar{\psi} \gamma_\mu \psi (0) 
= \int dt\, d\bar{t}\, \widetilde{C}^{A0}(t,\bar{t}\,) 
\, J_\mu^{A0}(t,\bar{t}\,)
\label{eq:A0current}
\end{equation}
with 
\begin{eqnarray}
&& J_\mu^{A0}(t,\bar{t}\,) = 
\bar{\chi}_{\bar{c}}(\bar{t} \nm)\gamma_{\perp\mu} \chi_c(t \np )\,,
\\
&& C^{A0}(\np p, \nm \bar{p}) = \int dt\, d\bar{t}\, 
e^{-i t \np p  -i \bar{t} \nm \bar{p}} \, \widetilde{C}^{A0}(t,\bar{t}\,) \,.
\end{eqnarray}
Here $\chi_c$ and $\mathcal{A}_{c\perp}^\mu$ (below) denote the 
collinear-gauge-invariant collinear quark and gluon fields that form the 
building blocks of general $N$-jet operators as defined in \cite{Beneke:2017ztn}.
Eq.~\eqref{eq:NLPfact} then turns into an intermediate result derived 
in \cite{Becher:2007ty}, which after further kinematic simplifications 
valid at LP leads to the LP factorization formula \eqref{eq:LPfact}.

We provide some further details on the definition of the generalized 
soft and the collinear functions. We define the soft Wilson lines
\begin{equation}
Y_{\pm}\left(x\right)=\mathbf{P}
\exp\left[ig_s\int_{-\infty}^{0}ds\,n_{\mp}A_{s}\left(x+sn_{\mp}\right)\right]
\end{equation}
in the fundamental representation, 
and perform the standard soft-decoupling transformation from (anti)collinear 
fields with $Y_+$ ($Y_-$). In terms of the decoupled collinear fields, 
which will be used below, the current reads
\begin{equation}
J_\mu^{A0}(t,\bar{t}\,) = 
\bar{\chi}_{\bar{c}}(\bar{t} \nm)Y^\dagger_-(0)
\gamma_{\perp\mu} Y_+(0)\chi_c(t \np )\,.
\label{eq:A0decoupled}
\end{equation}
We then introduce two sets of soft gluon and 
soft quark building blocks 
\begin{eqnarray}
\mathcal{B}_{\pm}^{\mu}&=&Y_{\pm}^{\dagger}\left[iD_{s}^{\mu}Y_{\pm}\right]\,,
\\
q^{\pm}&=&Y_{\pm}^{\dagger}q_{s}\,,
\end{eqnarray}
in terms of which the NLP soft-collinear SCET quark-gluon interaction 
Lagrangian \cite{Beneke:2002ni} can be written as 
\begin{eqnarray}
\mathcal{L}_{\xi}^{\left(1\right)}	
&=&\bar{\chi}_c ix_{\perp}^{\mu}
\left[in_{-}\partial\mathcal{B}_{\mu}^{+}\right]
\frac{\slashed n_{+}}{2}\chi_c 
\nonumber\\
\mathcal{L}_{1\xi}^{\left(2\right)} 	
&=&\frac{1}{2}\bar{\chi}_c i\nm x\,\np^{\mu}
\left[in_{-}\partial\mathcal{B}_{\mu}^{+}\right]
\frac{\slashed n_{+}}{2}\chi_c 
\nonumber\\
\mathcal{L}_{2\xi}^{\left(2\right)}	
&=&\frac{1}{2}\bar{\chi}_c x_{\perp}^{\mu}x_{\perp}^{\nu}
\left[i\partial_{\nu}in_{-}\partial\mathcal{B}_{\mu}^{+}\right]
\frac{\slashed n_{+}}{2}\chi_c 
\nonumber\\
\mathcal{L}_{3\xi}^{\left(2\right)}
&=&\frac{1}{2}\bar{\chi}_c x_{\perp}^{\mu}x_{\perp}^{\nu}
\left[\mathcal{B}_{\nu}^{+},in_{-}\partial\mathcal{B}_{\mu}^{+}\right]
\frac{\slashed n_{+}}{2}\chi_c 
\nonumber\\
\mathcal{L}_{4\xi}^{\left(2\right)}	
&=&\frac{1}{2}\bar{\chi}_c\left(i\slashed\partial_{\perp}+\sA_{c\perp}\right)
\frac{1}{i\np\partial}ix_{\perp}^{\mu}\gamma_{\perp}^{\nu}
\left[i\partial_{\nu}\mathcal{B}_{\mu}^{+}-i\partial_{\mu}\mathcal{B}_{\nu}^{+}
\right]\frac{\slashed n_{+}}{2}\chi_c + {\rm h.c.}
\nonumber\\
\mathcal{L}_{5\xi}^{\left(2\right)}	
&=&\frac{1}{2}\bar{\chi}_c\left(i\slashed\partial_{\perp}+\sA_{c\perp} \right)
\frac{1}{i\np\partial}ix_{\perp}^{\mu}\gamma_{\perp}^{\nu}
\left[\mathcal{B}_{\nu}^{+},\mathcal{B}_{\mu}^{+}\right]
\frac{\slashed n_{+}}{2}\chi_c+ {\rm h.c.}\nonumber\\
\mathcal{L}_{\xi q}^{\left(1\right)} 
&=&\bar{q}_{+}\sA_{c\perp}\chi_c+ {\rm h.c.}
\label{eq:scetinteractionsquark}
\end{eqnarray}
The soft-collinear interactions in the SCET Yang-Mills 
Lagrangian can be rewritten in a similar way.
We recall that the soft fields $\mathcal{B}_\mu^+$, $q^+$ are evaluated
at the multipole-expanded position. A corresponding expression holds 
for the anticollinear sector. The generalized soft functions are the vacuum 
matrix elements of Wilson lines from the DY current similar to
\eqref{eq:LPsoftfn} and in addition collections of 
$\mathcal{B}_\pm^\mu(z_\mp)$ 
insertions, where the argument $z_\mp^\mu = (n_\pm z)n_\mp^\mu/2$ 
arises from the integration 
over $d^4z$ in the time-ordered product and the multipole expansion 
of soft fields in soft-collinear interactions. Similar multilocal 
soft functions have appeared as ``subleading shape functions'' in 
early applications of 
SCET to power-suppressed effects in semi-inclusive heavy-quark 
decay \cite{Beneke:2004in,Lee:2004ja,Bosch:2004cb}, and 
recently in the resummation of NLP LLs of the 
thrust distribution in Higgs decay to two gluons \cite{Moult:2018jjd}.

A novel feature of the NLP factorization formula for the DY 
process is the appearance of collinear functions at the amplitude 
level~\cite{BBJVunpublished}. They are defined as the matching 
coefficients of a product of generic collinear fields to 
collinear-PDF fields in the presence of soft field operators. 
Equivalent definitions hold for the anticollinear sector. 
To this end, we construct a basis of gauge-covariant soft field
operators 
\begin{equation}
\mathfrak{s}_i(z_-)\in \left\{
\frac{\partial^{\mu}_{\perp}}{in_-\partial}
\mathcal{B}^+_{\mu_\perp}(z_-),
\partial_{[\mu_\perp}
\mathcal{B}^+_{\nu_\perp]}(z_-),
[\mathcal{B}^+_{\mu_\perp}(z_-)
,\,\mathcal{B}^+_{\nu_\perp}(z_-)],\ldots
\right\}
\end{equation}
from all the possible insertions of the subleading Lagrangian. 
We then collect the remaining collinear fields, and define the 
collinear functions $\widetilde{J}_{i}$ as the matching 
coefficients in 
equations, such as
\begin{eqnarray}
&&i\int d^4z\,
\mathbf{T}\left[ \chi_{c,\alpha a}(t n_+) \,
\mathcal{L}^{(2)}(z)\right] \nonumber\\
&& \hspace*{1.2cm}
= 2\pi\sum_i \int du \int \frac{d(n_+z)}{2} 
\,\widetilde{J}_{i;\alpha\beta,\mu,abd}\left(t,u;\frac{n_+z}{2}\right)
\, \chi^{\rm PDF}_{c,\beta b}(u n_+) \, 
\mathfrak{s}_{i;\mu,d}(z_-)\,,\qquad
\label{eq:collfn}
\end{eqnarray}
where $ab$ refer to colour and $\alpha\beta$ to Dirac indices, 
and $d$ ($\mu$) represents a collective colour (Lorentz) 
index from the soft operator.
In the DY process the c-PDF modes {\it can} be radiated into the final 
state without violating threshold kinematics, since their transverse 
momentum of order $\Lambda$ is negligible, hence the amplitude 
involves the matrix element 
$\langle X_{c,{\rm PDF}}|\ldots|A(p_A)\rangle$ of the above equation. 
After squaring the amplitude and summing over the unobserved
$X_{c,{\rm PDF}}$ particles along the beam direction, the matrix element 
$\langle X_{c,{\rm PDF}}|\chi^{\rm PDF}_{c,\beta b}(u n_+)|A(p_A)\rangle$ 
can be associated with the parton distribution function $f_{a/A}(x_a)$. 
The matching coefficient $\widetilde{J}_{i;\alpha\beta,\mu,abd}
(u,t;\omega)$ is a perturbative short-distance coefficient dominated 
by virtualities $Q^2(1-z)\gg \Lambda^2$. Since \eqref{eq:collfn} is 
an operator equation, the short-distance coefficient can be 
extracted by computing the partonic matrix element 
$\langle 0 |\ldots|q(x_ap_A)\rangle$. These considerations generalize 
to all collinear matrix elements that appear upon working out the 
time-ordered products with the subleading SCET Lagrangian. At leading 
twist in the $\Lambda/Q$ expansion, but to all orders in the 
$(1-z)$ expansion, only c-PDF operators with a single quark or a 
single gluon building block on the right-hand side of~\eqref{eq:collfn} 
are needed.  
Eq.~\eqref{eq:collfn} provides the SCET analogue and operator definition of 
the concept of `radiative jet functions' in the diagrammatic approach  
\cite{DelDuca:1990gz,Bonocore:2015esa,Bonocore:2016awd}. 

At NLP, the first subleading power in the $(1-z)$ expansion, the 
hard DY vertex should in principle be matched to SCET current 
operators with up to three collinear building blocks. The general 
basis of these operators and their renormalization is discussed in 
\cite{Beneke:2017ztn,Beneke:2018rbh}. In the 
classification of these papers the operators in question correspond 
to A0, A1, B1 and A2, B2, C2-type operators at orders 
$1,\lambda, \lambda^2$, 
respectively. However, a non-scaleless collinear function 
can arise only in conjunction with a time-ordered product, from 
the subleading SCET Lagrangian. Since the subleading Lagrangian 
insertions start at $\mathcal{O}(\lambda)$, the  
$\mathcal{O}(\lambda^2)$ suppressed A2, B2, C2-type can be 
dropped from the beginning. Further simplifications can be made 
when one is interested only in the leading logarithms, as described 
in the next section.

\section{Leading-logarithmic resummation}
\label{sec:LLresum}

\subsection{Relevant terms}
For the following discussion, we adopt a frame where the transverse 
momenta of the colliding partons vanish, 
$p_a^\mu = x_a \sqrt{s} n_-^\mu/2$,  $p_b^\mu = x_b
\sqrt{s} n_+^\mu/2$, and write the NLP factorization formula 
in the schematic form 
\begin{equation}
\hat\sigma  = [C\otimes J\otimes \bar{J}\,]^2 
\otimes S\,.
\label{eq:factsketch}
\end{equation}
Every factor including $\hat\sigma$ on the left-hand side depends 
on the renormalization and factorization scale $\mu$, but the 
scale dependence cancels after convolution with the parton densities
within a given accuracy, if all factors 
are computed consistently. Let us assume that $\mu$ 
is chosen at the collinear scale  $\mathcal{O}(Q\lambda)$. 
Since each factor depends only on 
a single scale, with this choice, the collinear functions 
$J$ do not contain large logarithms at any order in 
the expansion in the strong coupling $\alpha_s$. The hard ($C$) and soft 
($S$) functions exhibit large logarithms of $(1-z)$, which we 
aim to resum at LL.

The NLP LL series is given by the terms $\alpha_s^n \ln^{2n-1}(1-z)$. 
The soft functions start at $\mathcal{O}(\alpha_s)$ or higher, 
since they involve at least one soft gluon radiated into the final 
state. A NLP LL can be generated at one-loop only if the one-loop soft 
function contains $\alpha_s\ln (1-z)$ {\em and} if the product 
$[C\otimes J\otimes \bar{J}\,]^2$ has an 
$\mathcal{O}(\alpha_s^0)$ term. We can therefore drop all terms 
for which the hard and collinear functions do not have tree-level 
contributions.

Since at least one power of $\lambda$ comes with the necessary 
time-ordered product, at NLP $\mathcal{O}(\lambda^2)$, we can 
allow for at most one further factor $\lambda$ from the hard matching 
of the DY current. Hence, only one or none of the two factors of 
$C$ in \eqref{eq:factsketch} can be the coefficient function 
of an $\mathcal{O}(\lambda)$ suppressed current. The available 
structures are
\begin{eqnarray}
&& 
\bar{\chi}_{\bar c}(\bar{t}n_-)[n_\pm^\mu i\slashed{\partial}_{\perp}] 
\chi_{c}(t n_+), \;
\bar{\chi}_{\bar c}(\bar{t}n_-)[n_\pm^\mu (-i)
\overleftarrow{\slashed{\partial}}_{\!\perp}] 
\chi_{c}(t n_+) \quad (\mbox{A1-type})\,,
\nn\\
&& \bar{\chi}_{\bar c}(\bar{t}n_-)[n_\pm^\mu 
\slash{\mathcal{A}}_{c\perp}(t_2 n_+)] 
\chi_{c}(t_1 n_+), \;
\bar{\chi}_{\bar c}(\bar{t}_1n_-)[n_\pm^\mu 
\slash{\mathcal{A}}_{\bar{c}\perp}(\bar{t}_2 n_-)] 
\chi_{c}(t n_+)
\quad (\mbox{B1-type}).\;\;\qquad
\end{eqnarray}
The collinear gauge field in the B1-type currents must be contracted 
within the amplitude, which leads to collinear functions that start 
at the one-loop order $\mathcal{O}(\alpha_s)$. We can therefore 
neglect these operators. The A1-operators could produce NLP LL terms 
together with a soft function generated by a single 
$\mathcal{L}^{(1)}_\xi$ insertion. However, the 
time-ordered product of an A1-operator with $\mathcal{L}^{(1)}_\xi$ 
contains two collinear transverse indices from the current and 
$x_\perp^\mu$ and one soft transverse index from $\mathcal{B}_\mu^\pm$ 
in $\mathcal{L}^{(1)}_\xi$, whereas the leading-power SCET soft gluon 
interaction does not involve the transverse direction. The time-ordered 
product must then vanish in the adopted reference frame, 
since there is no external vector available to 
contract an odd number of transverse indices. We conclude that 
the NLP LL series arises entirely from the time-ordered products 
of subleading power soft-collinear 
Lagrangian interactions with the {\it leading-power}\/ SCET 
current \eqref{eq:A0current}. Moreover, its coefficient function 
can be taken in the tree-level approximation at the hard scale.

\subsubsection{Quark-antiquark channel}

The possible time-ordered products of the A0-operator 
at NLP can be inferred from \eqref{eq:scetinteractionsquark}
as well as the terms in ${\cal L}^{(1)}_{\rm YM}$ and 
 ${\cal L}^{(2)}_{\rm YM}$ 
in eqs.~(36) and (37) of \cite{Beneke:2002ni}.
We now argue that many of them do not contribute LLs.  

The single insertion of $\mathcal{L}^{(1)}_\xi$ or 
$\mathcal{L}^{(1)}_{{\rm YM}}$ vanishes, since the 
associated collinear function depends on a single transverse index, but 
an external transverse vector is not available in the chosen reference
frame. As expected, there is no $\mathcal{O}(\sqrt{1-z})$ power 
correction. The same argument is also valid for the 
case of two insertions of $\mathcal{L}^{(1)}_\xi$ or 
$\mathcal{L}^{(1)}_{{\rm YM}}$ arising from the
square of the amplitude with a single insertion.

We turn now to the single insertions of the 
$\mathcal{O}(\lambda^2)$ interactions 
in~\eqref{eq:scetinteractionsquark}.
The first step, as discussed before \eqref{eq:collfn}, 
is to find a minimal basis of soft fields operators. 
This is done systematically by considering Lagrangian insertions with 
one soft gluon, two soft gluons, and so on. For LL resummation only 
contributions from operators with a single soft gluon are necessary. 
This is because the soft function associated with the insertion
containing two soft gluons starts at $\mathcal{O}(\alpha_s^2)$.
Such soft functions could contribute to the NLP LL series
only if the $\mathcal{O}(\alpha_s^2)$ contribution has a
$\ln^3(1-z)$ term, which could arise, if there is logarithmically
enhanced one-loop mixing with a one-gluon soft 
function. No such logarithmic enhancements of the cusp anomalous 
dimension type are known for off-diagonal operator mixing and we 
assume the absence of such logarithmically enhanced mixing in 
the following. 
This argument also excludes the possibility of LL terms 
arising from a double insertion of $\mathcal{L}^{(1)}_\xi$
and the double insertion of $\mathcal{L}^{(1)}_{\xi q}$ 
into the amplitude as the corresponding soft functions also start at 
$\mathcal{O}(\alpha_s^2)$. The square 
of the single insertion of latter interaction contributes only to the 
quark-gluon channel. 

We restrict then to the $\mathcal{O}(\lambda^2)$ 
Lagrangian insertion with one soft gluon field. Inspecting 
\eqref{eq:scetinteractionsquark} and the YM Lagrangian we note that 
the possible soft gluon structures are $n_+\mathcal{B}^+$,
$\partial_{\nu} \mathcal{B}^+_{\perp \mu}$ for the collinear 
direction. The soft structure $n_+\mathcal{B}^+$, contained for 
example in ${\cal L}^{(2)}_{1\xi}$, can in fact be related to 
$\partial^{\mu}_{\perp}\mathcal{B}^+_{\perp \mu}$ 
using the equation of motion for the soft field, namely 
\be
\label{eq:eom}
n_+ {\cal B}^+ = -2 \frac{i \partial_{\perp}^{\mu}}
{i n_- \partial}{\cal B}_{\perp \mu}^+ 
+ \mbox{two-gluon terms.}
\ee
Furthermore, the  building block 
$\partial_{\perp\nu}\mathcal{B}^{+}_{\perp\mu}$ 
can be decomposed into $\frac{1}{2}
\partial_{\perp\{\nu} \mathcal{B}^+_{\perp \mu\}}+\frac{1}{2}
\partial_{\perp[\nu} \mathcal{B}^+_{\perp \mu]}$. The first, 
symmetric part must be proportional to $g^{\mu\nu}_{\perp}$, 
because the soft function is a vacuum matrix element of Wilson lines 
and the soft gluon field insertions. $g^{\mu\nu}_{\perp}$ is then 
the only symmetric structure which can carry two transverse Lorentz 
indices. The only remaining soft structure is given by 
the second, 
antisymmetric combination $i\partial_{\perp\nu}\mathcal{B}_{\perp\mu}^{+}
-i\partial_{\perp\mu}\mathcal{B}_{\perp\nu}^{+}$. This combination does 
not contribute, because by the above argument the NLP soft function 
with this soft field insertion must be proportional to the epsilon 
tensor, which is excluded by parity conservation of QCD.

We are thus left with only one collinear function proportional 
to a single soft building block, given by
\begin{eqnarray}   
&& i\int d^4z\,
\mathbf{T}\left[ \chi_{c,\alpha a}(t n_+) \Big(
\mathcal{L}^{(2)}_{1\xi}(z)+\mathcal{L}^{(2)}_{2\xi}(z)+
\mathcal{L}^{(2)}_{{\rm YM}}(z) \Big)\right] 
\nonumber \\ 
&& \hspace{1cm}
=\, 2\pi \int du \int \frac{d(n_+z)}{2} 
\,\widetilde{J}_{2\xi;\alpha\beta,abde}\left(t,u;\frac{n_+z}{2}\right)
\, \chi^{\rm PDF}_{c,\beta b}(u n_+) \, \frac{
\partial^{\mu}_{\perp}}{in_-\partial}
\mathcal{B}^+_{\perp\mu; de}(z_-)\,. \;\;
\quad
\label{collfn-1dot1}
\end{eqnarray}
We note at this point that $\mathcal{L}_{1\xi}^{(2)}$ contains 
$n_- z$, which in momentum space can be converted into a 
derivative $\partial_{n_+ p}$ on $C_{A0}(n_+p, n_- \bar{p})$. 
But at tree-level the hard coefficient is momentum-independent, 
and the derivative evaluates to zero. Moreover, the insertions of the
YM Lagranian can only start contributing through one loop.  
As discussed in section 3.1, only the tree level collinear function 
contribution is necessary for LL resummation, when $\mu$ is chosen 
at the collinear scale. Hence, the only Lagrangian insertion 
contributing at LL accuracy is given by 
$\mathcal{L}_{2\xi}^{(2)}$, since it is the only piece which 
contributes at tree level to the collinear function 
in (\ref{collfn-1dot1}). 

We therefore find that for LL resummation at NLP in the 
quark-antiquark channel only the single time-ordered product 
\begin{eqnarray}
\label{eq:Tproduct}
\left(J^{T2}_{A0,2\xi}(s,t)\right)^\mu = i \int d^4z \,
\mathbf{T}\left[  J^\mu_{A0}(s,t) \, \mathcal{L}^{(2)}_{2\xi}(z)\right]
\end{eqnarray}
needs to be considered when the renormalization scale 
$\mu$ is chosen 
at the collinear scale. To NLP LL accuracy the matching 
equation \eqref{eq:A0current} is then simply extended to 
\begin{equation}
\bar{\psi} \gamma^\mu \psi (0) 
= \int dt\, d\bar{t}\, \widetilde{C}^{A0}(t,\bar{t}\,) 
\left[
J^\mu_{A0}(t,\bar{t}\,) + 
\left(J^{T2}_{A0,2\xi}(t,\bar{t}\,)\right)^\mu 
+ \mbox{ $\bar{c}$-term}\right].
\label{eq:qqbarNLPamp}
\end{equation}
We shall consider explicitly the insertion on the incoming 
collinear quark line. There is the corresponding term, where 
the insertion is placed on the anticollinear antiquark line.
The NLP factorization formula \eqref{eq:NLPfact} for the 
$q\bar{q}$ channel  
is obtained after extracting the collinear function from the 
above equation, followed by squaring the amplitude. One also 
needs to keep track of a kinematic correction, which arises 
from evaluating the $d^3\vec{q}$, $d^3\vec{x}$ integrals in 
\eqref{eq:NLPfact} with NLP accuracy. 

\subsubsection{(Anti)quark-gluon channel}

For $g\bar{q} \to \gamma^* (\to\bar{\ell}\ell) + \bar{q}$ to occur near 
threshold, the PDF-collinear gluon needs to be converted into a 
collinear quark carrying almost all its momentum by emission of 
a soft antiquark. This process vanishes at LP, but can be realized at 
NLP by inserting $\mathcal{L}_{\xi q}^{(1)}$ once at the amplitude 
level. The matching equation for the (anti)quark gluon channel 
is 
\begin{equation}
\bar{\psi} \gamma^\mu \psi (0) 
= \int dt\, d\bar{t}\, \widetilde{C}^{A0}(t,\bar{t}\,) \left[
\left(J^{T1}_{A0,\xi q}(t,\bar{t}\,)\right)^\mu 
+ \mbox{ $\bar{c}$-term}\right]
\label{eq:gqNLPamp}
\end{equation}
with
\begin{eqnarray}
\label{eq:Tproductqg}
\left(J^{T1}_{A0,\xi q}(s,t)\right)^\mu = i \int d^4z \,
\mathbf{T}\left[  J^\mu_{A0}(s,t) \, 
\mathcal{L}^{(1)}_{\xi q}(z)\right]\,.
\end{eqnarray}
The cross section follows from the interference of this amplitude 
with itself. A non-vanishing interference requires that the two
$\mathcal{L}_{\xi q}^{(1)}$ insertions are either both on the 
collinear quark line, or both on the anticollinear antiquark line. 
No kinematic corrections need to be considered in this channel, 
since there is no LP amplitude.

\subsubsection{Gluon-gluon channel}

The gluon-gluon channel at threshold does not contain NLP leading 
logarithms for the production of a lepton-antilepton pair through 
a virtual photon that couples only to quarks. As shown above, 
NLP LLs could come only from the A0 quark-antiquark 
SCET current, but to turn the external PDF-(anti)collinear 
gluons into (anti)collinear (anti)quarks, would require at least 
four insertions of the $\mathcal{L}_{\xi q}^{(1)} $ Lagrangian, 
implying NNLP $\mathcal{O}(\lambda^4) \sim (1-z)^2$ suppression 
relative to the LP $q\bar{q}$ partonic channel. 
We do not discuss here the case of 
Higgs production for which this channel is relevant for the 
leading NLP logarithms.

\vskip0.2cm\noindent
In the following we focus on the quark-antiquark production channel 
and defer the further discussion of the quark-gluon channel to future work.

\subsection{Collinear functions}
The generic collinear modes with virtuality $Q^2\lambda^2$ are 
integrated out at the amplitude level by matching collinear field 
operators to PDF-collinear fields. In this section we define the 
collinear functions relevant to NLP LL resummation and 
calculate the tree-level coefficient functions.

So far we have shown that the only time-ordered product 
with insertion of ${\cal L}^{(2)}_{2\xi}$ contributes to 
LLs when $\mu$ is chosen at the collinear scale.
We collect the collinear fields from the time-ordered product 
\eqref{eq:Tproduct} in the operator
\begin{eqnarray}
\widetilde{\mathcal{J}}_{2\xi;\alpha,ade}(t n_+,z) = 
\frac{1}{2} \frac{z_{\perp}^2}{d-2}\,
(in_-\partial_z)^2 \,\mathbf{T}\left[\chi_{c,\alpha a}(tn_+)
\bar{\chi}_{c,d}(z) 
\frac{\slashed n_+}{2}\chi_{c,e}(z)\right]\,,
\end{eqnarray}
where $\alpha$ is a Dirac index, and Latin indices (except for $c$) 
are colour indices. The derivative prefactor $(in_-\partial_z)^2$ is 
conventional and conforms with the definition of the 
soft field product in \eqref{eq:softdef} below  
as well as being consistent with equation of motion in equation 
\eqref{eq:eom}.
We find it convenient to define 
the analogue of the matching equation \eqref{eq:collfn} 
in momentum space
with the soft field already stripped off.
Introducing  
\begin{eqnarray}
\label{2.6}
\mathcal{J}_{2\xi;\alpha,ade}(\np p,\omega) = 
\int dt\,e^{i (\np p) t}
\,i\int d^{4}z\,e^{i\omega (\np z)/2} \,
\widetilde{\mathcal{J}}_{2\xi;\alpha, ade}(tn_+,z)
\end{eqnarray}
and the Fourier-transform of the PDF-collinear 
quark field, 
\begin{eqnarray}
\hat{\chi}^{\text{\scriptsize PDF}}_{c,\alpha b}(\np p)
=\int du \,e^{i(\np p) u}
\,\chi^{\text{\scriptsize PDF}}_{c,\alpha b}(u n_{+})\,,
\label{eq:FTPDF}
\end{eqnarray}
the matching equation takes the form 
\begin{eqnarray}
\mathcal{J}_{2\xi;\alpha,ade}(\np p,\omega) = 
\int d(\np p^\prime) \, 
J_{2\xi;\alpha\beta,abde}(\np p,\np p^\prime;\omega)
\,\hat{\chi}^{\text{\scriptsize PDF}}_{c,\beta b}(\np p^\prime).
\end{eqnarray}
The collinear function 
$J_{2\xi;\alpha\beta,abde}(\np p,\np p^\prime;\, \omega)$ 
is defined as the matching coefficient in this operator equation. 

The matching coefficient is governed by the large scale $Q\lambda\gg 
\Lambda$. We determine it by calculating the $\langle 0|\ldots 
|q(p_a)\rangle$ matrix element of the above equation. At tree level
we find 
\begin{eqnarray}
J_{2\xi;\alpha\beta,abde}(\np p,\np p^\prime;\,\omega) &\equiv&  
J_{2\xi;\alpha\beta,abde}(\np p; \,\omega) \delta(\np p-\np p^\prime) 
\nonumber\\
&=& -\frac{1}{\np p}\delta(\np p-\np p^\prime) 
\delta_{\alpha\beta}\delta_{ad}\delta_{eb}\,.
\end{eqnarray}
For time-ordered products originating from A-type operators, which 
contain only a single collinear building block by definition, collinear 
momentum conservation implies that one can always extract a delta 
function, or derivatives of the delta function from the collinear 
function, as done above.  We can use the colour-Fierz relation
\begin{equation}
\delta_{ad}\delta_{eb} = \frac{1}{N_c}\delta_{ab}\delta_{ed} 
+ 2 \,T^A_{ab} T^A_{ed}
\end{equation}
with SU($N_c$) generators $T^A$ in the fundamental representation
to write 
\begin{eqnarray}
J_{2\xi;\alpha\beta,abde}(\np p;\, \omega) = 
J_{2\xi;\alpha\beta}^{(S)}(\np p;\, \omega)\delta_{ab}\delta_{ed}  + 
J_{2\xi;\alpha\beta}^{(O)}(\np p;\, \omega) 
T^A_{ab} T^A_{ed}
\end{eqnarray}
which is also the most general colour decomposition. Since from the 
definition the indices $de$ are contracted with the matrix soft  
gluon field $\mathcal{B}^\pm_{\perp\mu;de}$, the first term will never 
contribute. General arguments also imply that this collinear function 
is diagonal in the Dirac indices, hence to all orders 
in perturbation theory only a single scalar jet function 
defined through $J_{2\xi;\alpha\beta}^{(O)}(\np p;\omega) =  
\delta_{\alpha\beta} J_{2\xi}^{(O)}(\np p;\omega)$ arises from the time-ordered 
product~\eqref{eq:Tproduct}.

We now calculate the matrix element of \eqref{eq:qqbarNLPamp} 
after the (second) matching step that makes the collinear function 
explicit. The integrals over $t,\bar{t}$ can be performed by 
introducing the momentum-space hard function. After some 
manipulations, we obtain 
\begin{eqnarray}
\label{eq:NLPamp1}
&& \langle X|\bar{\psi} \gamma^\mu \psi (0) |A(p_A) B(p_B)\rangle 
=  
\int \frac{d\np p}{2\pi} \frac{d\nm \bar p}{2\pi}\,
C^{A0}(\np p,\nm \bar p)
\nonumber\\
&& \hspace*{0.5cm}\times
\int d\nm p_b\,\delta(\nm \bar{p} + \nm p_b)
\,\langle X_{\bar{c},\rm PDF}| 
\hat{\bar{\chi}}^{\rm PDF}_{\bar{c},\alpha a}
(\nm p_b)|B(p_B)\rangle
\nonumber\\
&& \hspace*{0.5cm}\times
\int d\np p_a \, \langle X_{c,\rm PDF}| 
\hat{\chi}^{\rm PDF}_{c,\beta b}(\np p_a)|A(p_A)\rangle
\nonumber\\
&& \hspace*{0.5cm}\times\, \bigg\{
\gamma^\mu_{\perp\alpha\beta}\,\delta(\np p - \np p_a)\,
\langle X_s|\mathbf{T}\!\left[Y^\dagger_-(0) Y_+(0)\right]_{ab}|0\rangle
\nonumber\\
&& \hspace*{1.2cm}+\,
\gamma^\mu_{\perp\alpha\gamma}\,\int \frac{d\omega}{4\pi}\,
J_{2\xi;\gamma\beta,fbde}\left(\np p,\np p_a;
\omega\right) 
\nonumber\\
&& \hspace*{1.2cm}\times 
\int d(\np z) \,e^{-i\omega (\np z)/2}\,
\langle X_s|\mathbf{T}\left(\left[Y^\dagger_-(0) Y_+(0)\right]_{af}
\frac{i\partial_{\perp\nu}}{i\nm\partial}
\mathcal{B}_{\perp\nu;de}^{+}(z_-)\right)|0\rangle
\nonumber\\
&& \hspace*{1.2cm}\,
\bigg\}+\,\mbox{$\bar{c}$-term}
\end{eqnarray} 
The above expression includes 
the LP term and the LP matching on the anticollinear antiquark leg. The 
new NLP contribution appears in the second- and third-to-last line. 
The corresponding power-suppressed contribution from the insertion on 
the anticollinear antiquark leg is the $\bar{c}$-term not written 
explicitly. 
Making use of the delta function in the collinear factor and its 
colour and spin decomposition, we can simplify 
\eqref{eq:NLPamp1} to
\begin{eqnarray}
\label{eq:NLPamp2}
&& \langle X|\bar{\psi} \gamma^\mu \psi (0) |A(p_A) B(p_B)\rangle 
=  \int \frac{d\np p_a}{2\pi} \frac{d\nm p_b}{2\pi}\,
C^{A0}(\np p_a,-\nm p_b)
\nonumber\\
&& \hspace*{0.5cm}\times
\,\langle X_{\bar{c},\rm PDF}| 
\hat{\bar{\chi}}^{\rm PDF}_{\bar{c},\alpha a}(\nm p_b)|B(p_B)
\rangle\,\gamma^\mu_{\perp\alpha\beta} \,\langle X_{c,\rm PDF}| 
\hat{\chi}^{\rm PDF}_{c,\beta b}(\np p_a)|A(p_A)\rangle\nonumber\\
&& \hspace*{0.5cm}\times\, \bigg\{
\,\langle X_s|\mathbf{T}\!\left[Y^\dagger_-(0) Y_+(0)\right]_{ab}|0\rangle
\nonumber\\
&& \hspace*{1.2cm}+\,
\frac{1}{2} \int \frac{d\omega}{4\pi}\,
J_{2\xi}^{(O)}(\np p_a;\omega)
\nonumber\\
&& \hspace*{1.2cm}\times
\int d(\np z) \,e^{-i\omega (\np z)/2}\,
\langle X_s|\mathbf{T}\left(\left[Y^\dagger_-(0) Y_+(0)\right]_{af}
\frac{i\partial_{\perp}^{\nu}}{i\nm\partial}
\mathcal{B}_{\perp\nu;fb}^{+}(z_-)\right)|0\rangle
\nonumber\\
&& \hspace*{1.2cm}\,
\bigg\}+\,\mbox{$\bar{c}$-term}
\end{eqnarray} 
Eq.~\eqref{eq:NLPamp1} as it stands is still valid to all orders in 
perturbation theory. In the tree approximation for the hard and 
collinear functions, which is sufficient to resum the NLP LLs, we can 
further set $C^{A0}(\np p_a,-\nm p_b)\to 1$ at the 
hard scale and 
$J_{2\xi}^{(O)}(\np p_a;\omega) \to -2/\np p_a$ at the collinear scale. 
The hadronic matrix 
elements of the PDF (anti)collinear state will turn into the 
(anti)quark parton distribution after squaring the above amplitude 
and summing over the hadronic final state $X$, which on the right-hand 
side has been split into its soft and PDF-(anti)collinear components.

\subsection{Soft functions}
\label{sec:soft}

The soft functions are defined after squaring the amplitude and 
summing over the soft final state $X_s$. 
We introduce the soft operator
\begin{equation}
\widetilde{\mathcal{S}}_{2\xi}\left(x,z_-\right)
= \mathbf{\bar{T}}\left[Y_{+}^{\dagger}(x)Y_{-}(x)\right]
\mathbf{T}\left[Y_{-}^{\dagger}(0)Y_{+}(0)
\frac{i\partial_{\perp}^{\nu}}{i\nm\partial}
\mathcal{B}_{\perp\nu}^{+}(z_-)\right]\,,
\label{eq:softdef}
\end{equation}
and the Fourier transform of its (colour-traced) vacuum matrix element
\begin{equation}
S_{2\xi}(\Omega,\omega) = 
\int\frac{dx^0}{4\pi}\int\frac{d(\np z)}{4\pi}\,
e^{i x^0\Omega/2- i\omega (\np z)/2} \,
\frac{1}{N_c}\,\mbox{Tr}\,
\langle 0 |\widetilde{\mathcal{S}}_{2\xi}(x^0,z_-)|0\rangle\,. 
\label{eq:s2xitraced}
\end{equation}

To see how this generalized soft function appears, we recall some 
standard steps in the derivation of the factorization formula for the 
DY process. We first express the phase-space delta function 
$(2\pi)^4 \delta(p_A + p_B - q -p_{X_s} - p_{X_{c,\rm PDF}} - 
p_{X_{\bar c,\rm PDF}})$ in terms of the space-time integral of 
the exponential, then use the translation operator to absorb the 
hadronic final state momenta into translations of the field 
arguments, then square the amplitude. At this point, the sum 
over the PDF-(anti)collinear 
state can be performed, and the matrix element of the 
PDF-(anti)collinear fields expressed in terms of the parton 
distributions, 
\begin{eqnarray}
&&\langle A(p_A)| \bar{\chi}_{c,\alpha a}(x+u^\prime\np)
\chi_{c,\beta b}(u\np) |A(p_A)\rangle = \frac{\delta_{ba}}{N_c} 
\left(\frac{\slashed{n}_-}{4}\right)_{\beta\alpha}\,
\np p_A
\nonumber\\
&& \hspace*{2cm} \times \int_0^1 dx_a\,f_{a/A}(x_a) \,
e^{i (x+u^\prime \np-u\np)\cdot x_ap_A}\,.
\end{eqnarray}
A similar standard definition applies to the anti-quark distribution 
in hadron $B$.\footnote{The contribution from antiquarks in $A$, 
and quarks in $B$ follows from a separate contribution with 
collinear and anti-collinear fields interchanged in the definition 
of the A0 current.} 

Applying these steps to \eqref{eq:NLPamp2}, performing the integrations 
over $\np p_a,\nm p_b$ and stripping off the convolution with the 
parton distribution functions, we arrive at
\begin{eqnarray}
\hat{\sigma}(z) &=& H(\hat{s})
\times \,Q^2 \int\frac{d^3\vec{q}}{(2\pi)^3 \,2\sqrt{Q^2+\vec{q}^{\,2}}}\,
\frac{1}{2\pi} \int d^4x\,
e^{i(x_a p_A+x_b p_B-q)\cdot x}
\nonumber\\
&&\times\left\{
\widetilde{S}_0(x) 
+ 2 \cdot \frac{1}{2}\int d\omega \,
J_{2\xi}^{(O)}(x_a \np p_A;\omega)\,
\widetilde{S}_{2\xi}(x,\omega) + \,\mbox{$\bar{c}$-term}\,
\right\}\,.\qquad
\label{eq:NLPfactqqbar}
\end{eqnarray}
Here $\widetilde{S}_0(x)$ is the leading-power position-space soft function, 
defined as the generalization of \eqref{eq:LPsoftfn} 
to $x^0\to x^\mu = (x^0,\vec{x}\,)$ in the position of the Wilson lines. 
The Fourier transform with respect to $x^0$ will be denoted 
by $S_0(\Omega,\vec{x})$ such that $S_{\rm DY}(\Omega) = S_0(\Omega,
\vec{0})$. Similarly $\widetilde{S}_{2\xi}(x,\omega)$ is the generalization 
of \eqref{eq:s2xitraced} such that $S_{2\xi}(\Omega,\omega)$ denotes 
the Fourier transform with respect to $x^0$ of $\widetilde{S}_{2\xi}
(x,\omega)_{|\vec{x}=0}$. The factor of two in \eqref{eq:NLPfactqqbar} 
arises from the two identical (see App.~\ref{app:softqqbar}) 
NLP terms in the square of the amplitude. We introduced the hard function
\begin{equation}
H(\hat{s},\mu_h) =  |C^{A0}(-\hat{s})|^2\,,
\end{equation}
which is the same for the LP and NLP term, and indicated the 
dependence on the hard renormalization scale for later convenience.

While the collinear function $J_{2\xi}^{(O)}(x_a \np p_A;\omega)$ is non-zero 
at tree level, the soft function starts only at $\order{\alpha_s}$. The 
straightforward one-loop calculation gives 
\begin{eqnarray}\label{eq:S2xi0}
S_{2\xi}(\Omega,\omega) = 
\frac{\alpha_s C_F}{2\pi}\left\{\theta(\Omega)\delta(\omega)
\left(-\frac{1}{\epsilon}+\ln\frac{\Omega^{2}}{\mu^{2}}\right)
+\left[\frac{1}{\omega}\right]_{+}\theta(\omega)
\theta(\Omega-\omega)\right\}\,,
\end{eqnarray}
where
\begin{equation}
\int_0^\Omega d\omega\, \frac{f(\omega)}{[\omega]_+}  
=\int_0^\Omega d\omega\, \frac{f(\omega)-f(0)}{\omega}\,.
\end{equation}
Eq.~\eqref{eq:S2xi0} exhibits a divergence despite being the lowest-order 
contribution. The divergence can be interpreted as mixing into the soft 
function $S_{x_0}$, defined as
\begin{eqnarray}
S_{x_0}(\Omega) = \int\frac{dx^0}{4\pi}\,e^{ix^0\Omega/2}
\,\frac{-2i}{x^0-i\varepsilon}\,\frac{1}{N_c}\,\mbox{Tr} \,\langle 0|
\mathbf{\bar{T}}\left[Y_+^{\dagger}(x^0)Y_-(x^0)\right]
\mathbf{T}\left[Y_-^{\dagger}(0)Y_+(0)\right]|0\rangle.\;
\quad
\label{eq:x0softfn}
\end{eqnarray}
In position space, this is simply the LP soft function with an 
extra factor $-2 i/x^0$, which leads to $\mathcal{O}(\lambda^2)$ 
power suppression and the presence of the $\theta(\Omega)$ factor 
in the tree-level expression $S_{x_0}(\Omega) = 
\theta(\Omega)$, required to cancel the divergence. At first sight, 
such soft functions might appear peculiar. However, similar objects 
with collinear fields were required in the renormalization of 
subleading gluon jet functions \cite{Becher:2010pd} and the above is 
the position-space and Drell-Yan process 
equivalent of the ``$\theta$-soft functions'' 
introduced in \cite{Moult:2018jjd} for the subleading-power 
thrust distribution.

We renormalize the soft functions by writing 
\begin{equation}
S_{A}(\Omega,\omega_i)_{|\textrm{ren}} = 
\sum_B \int d\Omega' \int d\omega'_j 
\, Z_{AB}(\Omega,\omega_i;\Omega',\omega'_j)
\,S_{B} (\Omega',\omega'_j)_{|\textrm{bare}}
\label{eq:softfnz}
\end{equation}
where $\omega_i, \omega'_j$ denote (possibly empty) sets of continuous 
variables that parameterize the non-locality of the soft function beyond the 
dependence on $\Omega$. In general, the number of arguments $\omega_{i}$ 
can be different than the number of $\omega_{j}'$, and the integration
is over all $\omega_{j}'$ that the bare soft function depends on.
If it depends only on $\Omega$, the integration $\int d\omega_{j}'$ 
can be omitted. Explicitly, the $2\xi$ soft function 
satisfies
\begin{eqnarray}
S_{2\xi} (\Omega,\omega)_{|\textrm{ren}} 
&=& \int d\Omega'\int d\omega'  
\, Z_{2\xi,2\xi}(\Omega,\omega;\Omega',\omega')
\,S_{2\xi} (\Omega',\omega')_{|\textrm{bare}}
\nonumber\\
&& +\,\int d\Omega'  \, Z_{2\xi,x_0}(\Omega,\omega;\Omega')
\,S_{x_0} (\Omega')_{|\textrm{bare}}
\label{eq:mix1}
\end{eqnarray}
with 
\begin{eqnarray}
&& Z_{2\xi,2\xi}(\Omega,\omega;\Omega,\omega') = 
\delta(\Omega-\Omega')\delta(\omega-\omega') + 
\mathcal{O}(\alpha_s)\,,
\\[0.1cm]
&& Z_{2\xi,x_0}(\Omega,\omega;\Omega') = 
\frac{\alpha_s C_F}{2\pi}\frac{1}{\epsilon}\delta(\Omega-\Omega')
\delta(\omega)+
\mathcal{O}(\alpha_s^2)\,.
\label{eq:z2xix0}
\end{eqnarray}
The mixing term in the second line of \eqref{eq:mix1} subtracts the 
divergent part of the first term on the right-hand side, resulting 
in a finite, renormalized soft function at $\mathcal{O}(\alpha_s)$.
The complete one-loop anomalous dimension matrix for the above soft 
functions required for LL resummation at NLP is derived in 
App.~\ref{app:softqqbar}.

\subsection{Kinematic corrections}

Kinematic corrections arise from the evaluation of the second line 
of \eqref{eq:NLPfact} with NLP accuracy. In the partonic center-of-mass 
frame $x_a \vec{p}_A + x_b \vec{p}_B  =0$, the three-momentum 
$\vec{p}_{X_s}$ of the soft hadronic final state is balanced by 
the lepton-pair, $\vec{q} + \vec{p}_{X_s}=0$. This implies the 
counting $\vec{q}\sim\lambda^2$, $q^0 = \sqrt{\hat{s}}+
\mathcal{O}(\lambda^2)$. The energy of the soft hadronic final state 
is expanded as 
\begin{equation}
[x_1 p_1+ x_2 p_2 -q]^0 =  p_{X_s}^0 = 
\sqrt{\hat{s}}-\sqrt{Q^2+\vec{q}^{\;2}} =
\frac{\Omega_*}{2} -\frac{\vec{q}^{\;2}}{2Q}+\order{\lambda^6}
\end{equation}
with
\begin{equation}
\Omega_* =  2 Q\frac{1-\sqrt{z}}{\sqrt{z}} = Q(1-z) + 
\frac{3}{4} Q(1-z)^2 +\order{\lambda^6}\,.
\label{eq:Omegastar}
\end{equation}
We then find for the second line 
of \eqref{eq:NLPfact} the approximation 
\begin{equation}
\frac{Q}{4\pi} \int dx^0\,e^{i x^0 \Omega_*/2}\left(1+\frac{i x^0
\partial_{\vec{x}}^2}{2 Q} +\order{\lambda^4}\right)
\widetilde{S}(x^0,\vec{x};\omega_i,\bar{\omega}_i,\omega_i^\prime,
\bar{\omega}_i^\prime)_{|\vec{x}=0}
\end{equation}
valid to NLP. It is understood that $\vec{x}=0$ is set after the spatial 
derivatives are taken. The result is general and holds for any soft function. 
However, at NLP we need to apply the kinematic correction only to the 
LP term. Eq.~\eqref{eq:NLPfactqqbar} therefore simplifies to 
\begin{eqnarray}
\label{eq:NLPfactqqbarsimp}
\hat{\sigma}(z) &=& H(\hat{s})
\times \,Q\,\bigg\{\left(1+
\frac{1}{Q}\frac{\partial}{\partial\Omega}\partial_{\vec{x}}^2\right)
S_0(\Omega,\vec{x})_{|\vec{x}=0,\Omega=\Omega_*}
\nonumber\\
&&+ \,2 \cdot \frac{1}{2}
\int d\omega \,J_{2\xi}^{(O)}(x_a\np p_A;\omega)\,
S_{2\xi}(Q(1-z),\omega) + \,\mbox{$\bar{c}$-term}
\,\bigg\} \,.
\end{eqnarray}
Since  $\widetilde{S}_0(x) = 1$ at tree level, the 
derivative soft function starts at $\mathcal{O}(\alpha_s)$. For dimensional 
reasons $x^0 \partial_{\vec{x}}^2 \widetilde{S}_0(x)_{|\vec{x}=0} 
\propto 1/x^0$. Hence, like the soft function $\widetilde{S}_{2\xi}$, 
the derivative soft function times $x^0$ mixes into $\widetilde{S}_{x_0}$ 
and produces leading logarithms at NLP. Further details on the 
renormalization of $\widetilde{S}_0(x)$ and 
$\partial_{\vec{x}}^2 \widetilde{S}_0(x)_{|\vec{x}=0}$ are provided in 
App.~\ref{app:softfnkin}.

The kinematic corrections contained in the first line of 
\eqref{eq:NLPfactqqbarsimp} can be made more explicit. First, 
additional corrections arise from expanding the hard matching coefficient 
$H(\hat s) = H(Q^2) + Q^2(1-z) H^\prime(Q^2) + \mathcal{O}(\lambda^4)$,
but these terms do not contribute to the NLP LL series. Indeed,   
$H^\prime(Q^2)$ starts with $\mathcal{O}(\alpha_s\ln(1-z))$, but 
the tree-level LP soft function is $\delta(1-z)$, which sets this 
term to zero. Any further term arising from the product of 
$(1-z) H^\prime(Q^2)$ with the LP soft function is explicitly at most of 
NLL accuracy. Hence we can set $H(\hat{s})\to H(Q^2)$ in 
\eqref{eq:NLPfactqqbarsimp}. Another implicit kinematic correction 
arises from the $(1-z)^2$ term in the definition \eqref{eq:Omegastar} 
of $\Omega_*$. Defining 
\begin{eqnarray}
S_{K1}(\Omega) &=& 
\frac{\partial}{\partial\Omega}\partial_{\vec{x}}^2 
S_0\left(\Omega,\vec{x}\right)_{|\vec{x}=0}\,,
\label{eq:sk1}\\
S_{K2}(\Omega) &=& \frac{3}{4}\,\Omega^{2}
\frac{\partial}{\partial\Omega}S_0(\Omega, 
\vec{x})_{|\vec{x}=0}\,,
\label{eq:sk2}\end{eqnarray}
the term in the first line of the curly bracket in 
\eqref{eq:NLPfactqqbarsimp} reads
\begin{equation}
S_{\rm DY}(Q(1-z))+\frac{1}{Q} S_{K1}(Q(1-z)) + \frac{1}{Q} S_{K2}(Q(1-z)) + 
\mathcal{O}(\lambda^4)\,.
\label{eq:kin2}
\end{equation}

Often, instead of the definition of the partonic cross section in 
\eqref{eq:dsigsq2}, the quantity
\begin{equation}
\Delta_{ab}(z) =\frac{\hat{\sigma}_{ab}(z)}{z}
\end{equation}
is considered. The additional NLP term $(1-z)\times \hat{\sigma}_{\rm LP}(z)$ 
in the expansion in $(1-z)$, 
where $\hat{\sigma}_{\rm LP}(z)$ refers to the LP term in the 
expansion of $\hat{\sigma}(z)$, 
can be formally included by adding the term $S_{K3}(Q(1-z))/Q$ 
to \eqref{eq:kin2}, where
\begin{eqnarray}
S_{K3}(\Omega)=\Omega\,S_0
(\Omega, \vec{x})_{|\vec{x}=0}.
\label{eq:sk3}
\end{eqnarray}
With this modification \eqref{eq:NLPfactqqbarsimp} applies to 
$\Delta(z)$ as well. 

All three soft functions $S_{Ki}(\Omega)$ vanish at 
$\mathcal{O}(\alpha_s^0)$, but mix into $S_{x_0}(\Omega)$ through 
a $1/\epsilon$ pole at the one-loop order. 
While each of the three functions is divergent and produces a NLP 
leading logarithm at $\mathcal{O}(\alpha_s)$, 
\begin{eqnarray}
S_{K1}\left(\Omega\right)&=&
\frac{\alpha_{s}C_{F}}{2\pi}\left( \frac{1}{\epsilon} 
+2\ln\frac{\mu}{\Omega}- 2\right)\theta\left(\Omega\right)
\\
S_{K2}\left(\Omega\right)&=&
\frac{\alpha_{s}C_{F}}{2\pi}\left( \frac{3}{\epsilon}
+6 \ln\frac{\mu}{\Omega}+6\right)\theta\left(\Omega\right)\\
S_{K3}\left(\Omega\right)&=&
\frac{\alpha_{s}C_{F}}{2\pi}\left(-\frac{4}{\epsilon}
-8 \ln\frac{\mu}{\Omega}\right)\theta\left(\Omega\right)\,,
\end{eqnarray}
the sum of all three kinematic corrections is finite, 
\begin{equation}
\sum_{i=1}^3 S_{Ki}(\Omega) = 2\,\frac{\alpha_s C_F}{\pi}
\,\theta(\Omega)
\end{equation}
and hence there is no leading logarithm. The diagonal 
renormalization of all three kinematic soft functions involves the same cusp 
anomalous dimension, since they descend 
from the same $S_0(x)$. The general structure of the renormalization 
group equation then implies that 
the cancellation of kinematic NLP leading logarithms for 
$\Delta_{q\bar{q}}(z)$ (but not $\hat{\sigma}_{q\bar q}(z)$) 
holds to all orders in perturbation theory, see App.~\ref{app:softfnkin}
for further details.

\subsection{Resummation}
\label{sec:resum}

We are now in the position to sum the leading logarithms at NLP to 
all orders in the $\alpha_s$ expansion. The logarithms arise from the 
ratio of the scales involved in the process. We shall sum the 
logarithms by evolving the hard function from the hard scale 
$\mu_h\sim Q$ and the soft functions from the soft 
scale $\mu_s\sim Q (1-z)$ to a common scale $\mu_c\sim Q\sqrt{1-z}$ 
using the renormalization group equations (RGEs) for the hard and 
soft functions. Choosing $\mu_c$ to be of order of the collinear scale, 
we do not need the RGE of the collinear function. We shall 
consider the expansion of $\Delta(z) = \hat{\sigma}(z)/z$, since, 
as discussed above, the kinematic corrections cancel for this 
quantity, which simplifies the discussion.

The hard matching function $H(Q^2,\mu)$ obeys the RGE 
\begin{equation}
\label{eq:RGE_CA0}
\frac{d}{d\ln\mu}H(Q^2,\mu)  
=\left(2\Gamma_{\text{cusp}}\ln\frac{Q^2}{\mu^2}+2 \gamma\right)
H(Q^2,\mu),
\end{equation}
which follows from the anomalous dimension of the LP SCET A0 
operator.\footnote{We note that the time-ordered products of 
Lagrangian insertions with the A0-operator do not mix into currents 
at LL accuracy \cite{Beneke:2018rbh}, which is consistent with the 
absence of power-suppressed current operators in the NLP factorization 
formula at LL accuracy.} 
Here 
\begin{equation}
\Gamma_{\text{cusp}}=\frac{\alpha_s}{\pi}C_{F}+\mathcal{O}(\alpha_s^2), 
\qquad
\gamma=-\frac{3}{2}\frac{\alpha_s}{\pi}C_{F}+\mathcal{O}(\alpha_s^2),
\label{eq:adhard}
\end{equation}
and $\alpha_s$ denotes the $\overline{\rm MS}$ QCD coupling at the 
scale $\mu$.
The general solution to \eqref{eq:RGE_CA0} reads
\begin{equation}
H(Q^2,\mu) = \exp\left[4 S(\mu_h,\mu) - 2 a_{\gamma}(\mu_h,\mu) 
\right] \left(\frac{Q^2}{\mu_h^2} \right)^{-2 a_\Gamma(\mu_h,\mu)}\!
H(Q^2,\mu_h) \,,
\label{eq:hardRGE}
\end{equation}
where \cite{Becher:2007ty}
\begin{eqnarray}
S(\nu,\mu) &=& - \int\limits_{\alpha_s(\nu)}^{\alpha_s(\mu)}\!
    d\alpha\,\frac{\Gamma_{\rm cusp}(\alpha)}{\beta(\alpha)}
    \int\limits_{\alpha_s(\nu)}^\alpha
    \frac{d\alpha'}{\beta(\alpha')}, 
\label{eq:Sdef}
\\ 
a_\Gamma(\nu,\mu) 
&=& - \int\limits_{\alpha_s(\nu)}^{\alpha_s(\mu)}\!
    d\alpha\,\frac{\Gamma_{\rm cusp}(\alpha)}{\beta(\alpha)} \,, \qquad  
a_\gamma(\nu,\mu) 
   = - \int\limits_{\alpha_s(\nu)}^{\alpha_s(\mu)}\!
    d\alpha\,\frac{\gamma(\alpha)}{\beta(\alpha)}.
\label{eq:adef}
\end{eqnarray}
To LL accuracy, $a_\Gamma$ and $a_\gamma$ can be set to zero, 
and $S(\nu,\mu)$ evaluated with the one-loop approximation to 
the cusp anomalous dimension and the beta function 
\begin{equation}
\beta(\alpha_s) = \frac{d}{d\ln\mu}\alpha_s = 
-2\,\frac{\beta_0\alpha_s^2}{4\pi} 
+\mathcal{O}(\alpha_s^3),
\qquad
\beta_0=\frac{11}{3} N_c-\frac{2}{3}n_f\,,
\end{equation}
resulting in 
\begin{eqnarray} 
S^{\rm LL}(\nu,\mu) &=&\frac{C_F}{\beta_0^2} 
\frac{4\pi}{\alpha_s(\nu)}
\left(1-\frac{\alpha_s(\nu)}{\alpha_s(\mu)}
+\ln \frac{\alpha_s(\nu)}{\alpha_s(\mu)}\right)\,.
\label{eq:SLL}
\end{eqnarray}

To evolve the soft function $S_{2\xi}$ we have to solve the coupled 
system of RGEs derived in \eqref{eq:ADM} of App.~\ref{app:softqqbar},
\begin{equation}
\frac{d}{d\ln\mu}\left(\begin{array}{c}
S_{2\xi}\left(\Omega,\omega\right)\\
S_{x_{0}}\left(\Omega\right)
\end{array}\right)=\frac{\alpha_{s}}{\pi}\left(\begin{array}{cc}
4C_{F}\ln\displaystyle\frac{\mu}{\mu_s} & -C_{F}\delta(\omega)\\
0 & 4C_{F}\ln\displaystyle\frac{\mu}{\mu_s}
\end{array}\right)\left(\begin{array}{c}
S_{2\xi}\left(\Omega,\omega\right)\\
S_{x^{0}}\left(\Omega\right)
\end{array}\right)\,,
\label{eq:ADMcopy}
\end{equation}
where $\mu_s$ denotes an arbitrary soft scale of 
order $Q(1-z)$. This equation matches precisely the form of 
the general equation \eqref{eq:rgeSNLP2by2}, if we identify 
$L_0 \to \ln(\mu^2/\mu_s^2)$, $\Delta^{(0)} = -C_F$. 
Applying \eqref{eq:LLsol}, \eqref{eq:LLfns} to this specific 
case, we can immediately write down the LL solution
  \begin{equation}
  S_{2\xi}^{\rm LL}(\Omega,\omega,\mu) =
  \frac{2 C_F}{\beta_0}
  \ln\frac{\alpha_s(\mu)}{\alpha_s(\mu_s)}\,
  \mbox{exp}\left[-4 S^{\rm LL}(\mu_s,\mu)\right]
  \,\theta(\Omega)\delta(\omega)\,.
  \label{eq:LLsol2xi}
  \end{equation}
Here $\mu$ does not have to be chosen of order of the soft scale, 
in which case the solution sums the leading large logarithms 
$\ln(\mu/\mu_s)$ to all orders.

With the LL evolved hard and soft function at hand, we can proceed 
with the evaluation of the NLP partonic cross section 
\eqref{eq:NLPfactqqbarsimp}. We write this equation for 
$\Delta(z) = \hat{\sigma}(z)/z$, in which case we can drop the 
kinematic correction at LL accuracy, hence \eqref{eq:NLPfactqqbarsimp} 
simplifies to 
\begin{eqnarray}
\label{eq:NLPfactqqbarsimpevolved}
\Delta(z) &=& H(Q^2,\mu_c)
\times \,Q\,\bigg\{
S_{\rm DY}(Q (1-z),\mu_c)
\nonumber\\
&&+ \,2 \cdot \frac{1}{2}
\int d\omega \,J_{2\xi}^{(O)}(x_a\np p_A;\omega,\mu_c)\,
S_{2\xi}(Q(1-z),\omega,\mu_c) + \,\mbox{$\bar{c}$-term}
\,\bigg\}\,.
\end{eqnarray}
We have explicitly indicated the scale dependence of all quantities and 
it is understood that $H$ and $S_{2\xi}$ are evolved from $\mu_h\sim 
Q$ and $\mu_s\sim Q(1-z)$ to a common scale $\mu_c \sim Q\sqrt{1-z}$. 
Inserting the tree-level value of the collinear function 
$J_{2\xi}^{(O)}(x_a\np p_A;\omega,\mu) = -2/(x_a \np p_A)$ and using 
$x_a \np p_A=Q + \mathcal{O}(Q(1-z))$ in the NLP term, we find 
\begin{eqnarray}
\Delta(z) &=& H(Q^2,\mu_c)
\times \,Q\,\bigg\{
S_{\rm DY}(Q (1-z),\mu_c)
\nonumber\\
&& 
- \,\frac{2}{Q}
\int d\omega \,
S_{2\xi}(Q(1-z),\omega,\mu_c) + \,\mbox{$\bar{c}$-term}
\,\bigg\}\,.
\label{eq:NLPfactevolved2}
\end{eqnarray}
At this point the $\bar{c}$-term takes an identical form to the 
second term in the curly bracket and the two can be added. In effect, 
the four insertions of the $\mathcal{L}^{(2)}_{2\xi}$ Lagrangian 
on the four external legs of the squared amplitude all contribute 
the same amount to the leading-logarithmic next-to-leading power 
correction. We refer to App.~\ref{app:softqqbar} for the details of the 
argument.  

With the explicit LL solutions \eqref{eq:hardRGE}, \eqref{eq:LLsol2xi} 
for $H(Q^2,\mu)$ and $S_{2\xi}(\Omega,\omega,\mu)$, respectively, 
inserted, the above equation reads
\begin{eqnarray}
\Delta^{\rm LL}(z) &=&\Delta^{\rm LL}_{\rm LP}(z)
\nonumber\\
&&  -\,
\mbox{exp}\left[4 S^{\rm LL}(\mu_h,\mu_c)-4 S^{\rm LL}(\mu_s,\mu_c)\right]
\times \frac{8C_F}{\beta_0}
\ln\frac{\alpha_s(\mu_c)}{\alpha_s(\mu_s)}\,\theta(1-z)\,,\quad
\label{eq:NLPsummedfinal}
\end{eqnarray}
where we also used $H(Q^2,\mu_h) = 1+\mathcal{O}(\alpha_s)$. 
$\Delta^{\rm LL}_{\rm LP}(z)$ represents the LL-resummed 
leading-power partonic cross section, in the present formalism 
given in \cite{Becher:2007ty}. 
We can set $\mu_h=Q$, $\mu_s=Q(1-z)$ and $\mu_c=Q\sqrt{1-z}$, since 
the precise choice is irrelevant for the leading logarithms. 

Note, however, that \eqref{eq:NLPsummedfinal} is not of the most general 
form, since it implies that the factorization scale $\mu$ is  
set to $\mu_c=Q\sqrt{1-z}$ in the parton distributions. 
In order to translate the result to arbitrary 
$\mu$ we use the evolution equation for the partonic cross section, 
\begin{equation}\label{eq:consistency}
\frac{d}{d\ln\mu}\hat\sigma_{ab}(z,\mu) = -
\sum_c \int_z^1dx
\left(P_{ca}(x)\hat\sigma_{cb}\left(\frac{z}{x},\mu\right)
+P_{cb}(x)\hat\sigma_{ac}\left(\frac{z}{x},\mu\right)\right).
\end{equation}
In order to extract the power-suppressed contribution at 
${\cal O}(\lambda^2)$, we expand the Altarelli-Parisi splitting kernels 
in the form 
\be
 P_{ab}(x)= P_{ab}^{\rm LP}(x) + P^{\rm NLP}_{ab} + {\cal O}(1-x)\,,
\ee
where
\bea\label{eq:Pabexpanded}
&&  P_{ab}^{\rm LP}(x)=\left(2\Gamma_{\rm cusp}(\alpha_s)\,\frac{1}{[1-x]_+} + 
2\gamma^\phi(\alpha_s)\delta(1-x) \right)\delta_{ab} \,,\nn\\
&& P_{ab}^{\rm NLP} =\gamma^{\rm NLP}_{ab}(\alpha_s)\,.
\eea
For $a=b=q$ or $\bar q$, $\gamma^\phi=\frac34\alpha_s C_F/\pi
+{\cal O}(\alpha_s^2)$ and 
$\gamma^{\rm NLP}_{qq}=\gamma^{\rm NLP}_{\bar q\bar q}=-2\alpha_s C_F/\pi
+\mathcal{O}(\alpha_s^2)$.
For the $q\bar q$ channel, no mixing terms in the splitting kernels 
need to be considered up to ${\cal O}(\lambda^2)$, because $P_{ab}$
starts at ${\cal O}(\lambda^2)$ for $a\not=b$, and the contributions from 
$\hat\sigma_{ab}$ for $ab\not=q\bar q$ yield additional power-suppression,
starting at $\mathcal{O}(\lambda^2)$ for the $qg$ channel.
For the quantity $\Delta(z)=\hat\sigma_{q\bar q}(z)/z$ this implies
\begin{equation}\label{eq:consistencyDelta}
\frac{d}{d\ln\mu}\Delta(z,\mu) = -2
 \int_z^1\frac{dx}{x}
P_{qq}(x)\Delta\left(\frac{z}{x},\mu\right)+{\cal O}(\lambda^4)\,,
\end{equation}
where we used also $P_{qq}=P_{\bar q\bar q}$.

In App.~\ref{app:derivation} we derive from this equation the 
evolution equation for the NLP part $\Delta_{\rm NLP}(z,\mu)$ in 
the expansion $\Delta(z,\mu)=\Delta_{\rm LP}(z,\mu)+
\Delta_{\rm NLP}(z,\mu)+\dots$, and find 
\bea\label{eq:rge_delta}
\lefteqn{\frac{d}{d\ln\mu}\Delta_{\rm NLP}(z,\mu) } \nn\\
&=&-4\left[\Gamma_{\rm cusp}(\alpha_s) \left(\ln(1-z)
-\gamma_E-\psi\left(1+\frac{d}{d\ln(1-z)}\right)\right)
+\gamma^\phi(\alpha_s)\right]\Delta_{\rm NLP}(z,\mu)\nn\\
&&   + \,K(z,\mu)\,,
\eea
with the inhomogeneous term given by
\be
\label{eq:Kdef}
K(z,\mu) =  -2\gamma^{\rm NLP}_{qq}(\alpha_s)
\int_z^1dy\,\Delta_{\rm LP}\left(y,\mu\right) 
-4\,\Gamma_{\rm cusp}(\alpha_s)(1-z)\Delta_{\rm LP}(z,\mu)\,.
\ee
Since we are interested in LL accuracy only,
all terms in the square bracket of \eqref{eq:rge_delta} 
except for $\Gamma_{\rm cusp}(\alpha_s)\ln(1-z)$
can be dropped, as well as the second term on the right-hand side of 
\eqref{eq:Kdef}, that arises from the kinematic correction.
The solution of \eqref{eq:rge_delta} under this approximation
is given by
\begin{eqnarray}
\Delta_{\rm NLP}(z,\mu) = 
e^{\hat S(z,\mu_c,\mu)}\, \Delta_{\rm NLP}(z,\mu_c) + 
\int_{\ln\mu_c}^{\ln\mu} d\ln\mu'\,
e^{\hat S(z,\mu',\mu)}\,K(z,\mu')\,,
\label{eq:Deltasol}
\end{eqnarray} 
with 
\begin{equation}
\hat{S}(z,\nu,\mu) = 4 a_\Gamma(\nu,\mu)\ln(1-z)\,.
\end{equation}

We use the LL approximation to this solution to evolve the 
NLP term in \eqref{eq:NLPsummedfinal} from $\mu_c$ to an 
arbitrary $\mu$. 
To calculate the inhomogeneous term to LL accuracy, we use the 
expression 
\begin{equation}
\Delta_{\rm LP}^{\rm LL}(z,\mu) = 
\mbox{exp}\left[4 S^{\rm LL}(\mu_h,\mu)-4 S^{\rm LL}(\mu_s,\mu)\right]
\frac{e^{-2\gamma_E\eta}}{\Gamma(2\eta)}
\frac{1}{1-z}\left(\frac{Q(1-z)}{\mu_s}\right)^{2\eta}\,,
\end{equation}
for the LL resummed LP partonic cross section, 
where $\eta = 2a_\Gamma(\mu_s,\mu)$ and the expression must be 
understood as a distribution \cite{Becher:2007ty}. Recalling that 
the expansion of $S^{\rm LL}$ yields double logarithms at every 
order, while that of $\eta$ produces only single logarithms, 
we obtain from \eqref{eq:Kdef}
\be
 K^{\rm LL}(z,\mu)=\frac{4\alpha_s(\mu) C_F}{\pi}
\mbox{exp}\left[4 S^{\rm LL}(\mu_h,\mu)-4 S^{\rm LL}(\mu_s,\mu)\right]
\theta(1-z)\,.
\label{eq:Ksol}
\ee
To leading-logarithmic accuracy 
\begin{equation} 
4 S^{\rm LL}(\mu_h,\mu)-4 S^{\rm LL}(\mu_s,\mu) = \hat S(z,\mu_c,\mu)\,, 
\end{equation}
which also implies $S^{\rm LL}(\mu_h,\mu_c)-S^{\rm LL}(\mu_s,\mu_c)=0$ 
in \eqref{eq:NLPsummedfinal}. Hence the two terms on the right-hand side 
of (\ref{eq:Deltasol}) are given by
\begin{eqnarray}
&& e^{\hat S(z,\mu_c,\mu)}\, \Delta_{\rm NLP}(z,\mu_c)  
= e^{\hat S(z,\mu_c,\mu)}\times  \frac{-8C_F}{\beta_0}
\ln\frac{\alpha_s(\mu_c)}{\alpha_s(\mu_s)}\,\theta(1-z)
\nonumber\\
&& \hspace*{1cm} 
=\, \mbox{exp}\left[4 S^{\rm LL}(\mu_h,\mu)-4 S^{\rm LL}(\mu_s,\mu)\right]
\times \frac{-8C_F}{\beta_0}
\ln\frac{\alpha_s(\mu_c)}{\alpha_s(\mu_s)}\,\theta(1-z)\,,\quad
\\[0.2cm]
&& \int_{\ln\mu_c}^{\ln\mu} d\ln\mu'\,
e^{\hat S(z,\mu',\mu)}\,K(z,\mu')\
= e^{\hat S(z,\mu_c,\mu)} \int_{\ln\mu_c}^{\ln\mu} d\ln\mu'\,
\frac{4\alpha_s(\mu')C_F}{\pi} 
\nonumber\\
&& \hspace*{1cm} 
=\, \mbox{exp}\left[4 S^{\rm LL}(\mu_h,\mu)-4 S^{\rm LL}(\mu_s,\mu)\right]
\times \frac{-8C_F}{\beta_0}
\ln\frac{\alpha_s(\mu)}{\alpha_s(\mu_c)}\,\theta(1-z)\,,
\end{eqnarray}
where we used $\hat{S}(z,\mu',\mu) + \hat{S}(z,\mu_c,\mu') =   
\hat{S}(z,\mu_c,\mu)$.  Adding the two terms we find
\begin{eqnarray}
\Delta^{\rm LL}_{\rm NLP}(z,\mu) &=& 
\mbox{exp}\left[4 S^{\rm LL}(\mu_h,\mu)-4 S^{\rm LL}(\mu_s,\mu)\right]
\times \frac{-8C_F}{\beta_0}
\ln\frac{\alpha_s(\mu)}{\alpha_s(\mu_s)}\,\theta(1-z)\,,\quad
\label{eq:NLPsummedfinalgeneral}
\end{eqnarray}
which is identical to the NLP term in \eqref{eq:NLPsummedfinal} except 
that $\mu$ in the above expression is not restricted to the collinear scale,
but can take any value. 
The scale of the parton luminosity that multiplies the above expression is 
now manifestly independent of $z$, and the logarithms of $(1-z)$ 
are generated by setting $\mu_s\sim Q(1-z)$. The fact that the form 
of \eqref{eq:NLPsummedfinal} and \eqref{eq:NLPsummedfinalgeneral} 
are identical implies that the collinear function cannot contain 
leading logarithms when evaluated at a scale $\mu$ different from 
$\mu_c$. 

The simple result \eqref{eq:NLPsummedfinalgeneral} 
for the summed NLP LLs is the main result of this paper. Let us briefly 
summarize the main steps of the derivation. Starting from the 
NLP factorization formula (\ref{eq:factsketch}), we obtained 
the explicit LL-accurate form (\ref{eq:NLPfactqqbarsimp}), which contains 
NLP terms from a kinematic correction and the convolution of a 
radiative jet function with a generalized soft function. We then 
noted that the kinematic correction cancels for the quantity $\Delta(z)$ 
and obtained (\ref{eq:NLPfactevolved2}). The LL resummed NLP correction 
(\ref{eq:NLPsummedfinal}) follows from inserting the LL resummed 
generalized soft function (\ref{eq:LLsol2xi}) and the well-known hard 
function. The final step to arrive at \eqref{eq:NLPsummedfinalgeneral} 
consisted in evolving (\ref{eq:NLPsummedfinal}) from the collinear scale 
$\mu_c$ to an arbitrary $\mu$, which turned out to preserve the 
form of the resummed result.

\section{Fixed-order expansion: predictions and
checks}
\label{sec:FOexpansion}

We expand the resummed NLP cross section in the strong coupling.
This serves both, as a check of the equation by comparing the logarithms
at fixed order with the known NLO and NNLO partonic cross sections
and provides new results beyond these orders.

To generate the fixed-order logarithms from \eqref{eq:NLPsummedfinalgeneral},
we expand the ratios of the running strong coupling into a series
in $\alpha_s(\mu)$ and logarithms. To LL accuracy, we may approximate
\begin{eqnarray}
S^{\rm LL}(\mu_1,\mu_2) = -\frac{\alpha_s C_F}{2\pi}
\ln^2\frac{\mu_2}{\mu_1}\,,
\qquad\quad
\frac{1}{\beta_0} \ln\frac{\alpha_s(\mu_1)}{\alpha_s(\mu_2)} =
\frac{\alpha_s}{2\pi} \ln\frac{\mu_2}{\mu_1}\,.
\end{eqnarray}
The scale of $\alpha_s$ on the right-hand sides of these equations
is not specified, since its precise value is a NLL effect. The NLP
term in \eqref{eq:NLPsummedfinalgeneral}, with $\mu$ taken to be the
free renormalization and factorization scale as discussed above, then 
reduces to
\begin{eqnarray}
\Delta^{\rm LL}_{\rm NLP}(z,\mu) =  
\frac{\hat\sigma^{\rm LL}_{\rm NLP}(z,\mu)}{z} &=&
\mbox{exp}\left[ -2 \frac{\alpha_s C_F}{\pi}
\ln^2\frac{\mu}{\mu_h} \right]\,
\mbox{exp}\left[+2 \frac{\alpha_s C_F}{\pi}
\ln^2\frac{\mu}{\mu_s}\right]\nn\\
&& \times (-4) \frac{\alpha_s C_F}{\pi}
\ln\frac{\mu_s}{\mu}\,\theta(1-z)\,.
\label{eq:NLPsummedfinallog}
\end{eqnarray}
Interestingly, for the special choice $\mu=\mu_c$, the two exponentials cancel
to LL accuracy and the NLP LL series becomes trivial,
\begin{eqnarray}
\Delta^{\rm LL}_{\rm NLP}(z,\mu_c) &=&
-2 \,\frac{\alpha_s C_F}{\pi} \ln(1-z)\,\theta(1-z)\,.
\end{eqnarray}

In this case, the $z$-dependence of the hadronic cross section is hidden 
in the scale-dependent parton densities, evaluated at the collinear scale.
For arbitrary $\mu$, we use \eqref{eq:NLPsummedfinallog} with 
$\mu_h=Q$ and $\mu_s=Q(1-z)$, and define 
$L_\mu=\ln(\mu/Q)$. Then 
\begin{eqnarray}
\Delta^{\rm LL}_{\rm NLP}(z,\mu) & =&
 {} - \theta(1-z)\,\bigg\{4C_F\frac{\alpha_s}{\pi}\Big[\ln(1-z)-
L_\mu \,\Big]\nn\\
&& \hspace*{-1.5cm} + \,8C_F^2\left(\frac{\alpha_s}{\pi}\right)^2
\Big[\ln^3(1-z)-3 L_\mu \ln^2(1-z)+2 L_\mu^2 \ln(1-z)\,\Big]
\nn\\
&&  \hspace*{-1.5cm}+ \,8C_F^3\left(\frac{\alpha_s}{\pi}\right)^3
\Big[\ln^5(1-z)-5 L_\mu \ln^4(1-z) +8 L_\mu^2 \ln^3(1-z) 
-4 L_\mu^3 \ln^2(1-z)\,\Big]
\nn\\
&&  \hspace*{-1.5cm}+ \,\frac{16}{3}C_F^4\left(\frac{\alpha_s}{\pi}\right)^4
\Big[\ln^7(1-z)-7 L_\mu\ln^6(1-z) +18 L_\mu^2 \ln^5(1-z) 
-20 L_\mu^3 \ln^4(1-z)\nn\\
&& + \, 8L_\mu^4 \ln^3(1-z)\,\Big]
\nn\\
&&  \hspace*{-1.5cm}+ \,
\frac{8}{3}C_F^5\left(\frac{\alpha_s}{\pi}\right)^5
\Big[\ln^9(1-z)-9 L_\mu\ln^8(1-z) +32 L_\mu^2 \ln^7(1-z) 
 -56L_\mu^3 \ln^6(1-z)\nn\\
&&
+\,48 L_\mu^4 \ln^5(1-z)-16L_\mu^5 \ln^4(1-z)
\,\Big]\bigg\} + \,{\cal O}(\alpha_s^6\times (\log)^{11})\,,
\label{eq:fullexpanded}
\end{eqnarray}
where $(\log)^{11}$ stands for some combination of the two logarithms to 
the 11th power.

The first two lines can be compared to known exact results 
\cite{Hamberg:1990np}.\footnote{Our $\Delta(z,\mu)$
 corresponds to $\Delta_{ij}(x,Q^2,M^2)$ in the
notation of \cite{Hamberg:1990np}, where $ij=q\bar q$, $x=z$, and $M=\mu$.}
In particular, the NLO result in the first line agrees with Eq.~(B.29) 
in \cite{Hamberg:1990np},
and the NNLO contribution in the second line with Eq.~(B.31) in 
\cite{Hamberg:1990np}, up to subleading terms in the expansion in 
logarithms.  The N$^3$LO term confirms the 
conjecture \cite{Kramer:1996iq,Kidonakis:2007ww} that 
the leading logarithm at this order can be simply obtained from 
including the NLP term in the Altarelli-Parisi splitting kernels 
in the standard LP resummation formalism. 
The N$^3$LO and N$^4$LO terms for $L_\mu=0$ have been given in 
Eqs.~(B.4) and (B.5) of \cite{deFlorian:2014vta} based on the 
observation that the ``physical evolution kernels'', which express 
the scale dependence of a given quantity in terms of the quantity 
itself, exhibits only single logarithms as $z\to 1$ to the 
respective order in the $\alpha_s$ expansion. Our direct derivation 
from the all-order resummed partonic cross section 
agrees with these results as well.
The N$^5$LO term is a new result and the 
expansion to any order can be obtained without effort from 
\eqref{eq:NLPsummedfinallog}.

\section{Summary}
\label{sec:summary}

Historically, soft gluon resummation was first studied for the threshold of 
the DY process \cite{Sterman:1986aj,Catani:1989ne} and then extended to increasingly higher 
logarithmic accuracy at leading power in the expansion in the threshold 
variable $(1-z)$, based on the factorization into a hard and soft function. 
In this work, we considered the next-to-leading power (NLP) in $(1-z)$
and provided an all-order resummation of the leading NLP logarithms 
of the form $\alpha_s^n \ln^{2 n-1}(1-z)$, $n=1,2,\ldots$ near the kinematic 
threshold $z=Q^2/\hat{s}\to 1$. 

This result is made possible by the systematic treatment of the 
threshold in the framework of position-space soft-collinear effective 
theory, which implies a generalized factorization formula including 
collinear functions at the amplitude level beyond leading power. We sketched 
and explained the factorization formula, which will be further elaborated in 
\cite{BBJVunpublished}, and derived a simple expression for the 
leading logarithmic terms at NLP in the quark-antiquark production 
channel. Our main results are the resummed partonic cross sections 
\eqref{eq:NLPsummedfinalgeneral} and \eqref{eq:NLPsummedfinallog}, which 
follow from the solution to the renormalization group equations for the 
hard function and generalized subleading-power soft functions. 
The final result is stunningly simple. When the parton distributions 
are evaluated at a factorization scale of order $Q\sqrt{1-z}$, the 
LL series terminates at $\mathcal{O}(\alpha_s)$.  For general 
factorization scale, re-expansion in $\alpha_s$ shows agreement with 
known exact results at NLO and NNLO and with predictions from the physical 
evolution kernel at N$^3$LO and N$^4$LO, and leads to new results 
at the five-loop order and beyond. 

We are aware of two other resummations at subleading power, both 
also to LL accuracy. The thrust distribution in Higgs decay to 
two gluons for  $(1-T)\to 1$ was recently considered, also in the 
SCET framework \cite{Moult:2018jjd}. There are differences 
of conceptual nature in the collinear physics, as thrust is 
infrared-safe, while the parton distributions must be factorized 
for the DY process, which requires the introduction of the 
PDF collinear modes and the matching of the collinear functions 
at the amplitude level. Nevertheless, at the LL level, the 
NLP resummation has a very similar structure and the result a similar 
level of simplicity. In particular, the mixing pattern of the 
required subleading soft functions is similar, and the ``theta soft-function'' 
(here the $1/x^0$ soft function) appears in both cases. A number of 
quark-mass suppressed form factors has been resummed with 
LL accuracy in \cite{Liu:2017vkm,Liu:2018czl} through a diagrammatic 
all-order analysis. Here the NLP suppression 
is provided by the quark mass $m\ll Q$ rather than the kinematics 
itself. It would be interesting to reproduce and generalize 
this result in the SCET framework.

All NLP resummations are presently restricted to the leading 
logarithmic accuracy. We expect extensions of these results to 
NLL to reveal the full complexity of NLP resummation. Indication 
of this is already provided by the form of the non-cusp 
terms of the one-loop anomalous 
dimension kernels of subleading-power $N$-jet operators 
\cite{Beneke:2017ztn,Beneke:2018rbh}, which enter at this order. 
In addition, the renormalization of subleading-power soft and 
collinear functions must be better understood to determine the 
single-pole terms needed for NLL resummation. A related question 
is whether the convolution integrals in the $\omega_i$ variables 
of collinear and soft functions exhibit singularities at 
$\omega_i\to 0$, requiring extra renormalization.

\subsubsection*{Acknowledgments} 
We thank T.~Becher and B.~Mistlberger for an interesting conversation. 
MB thanks the Albert Einstein Institute at Bern University and the 
Center for the Fundamental Laws of Nature at Harvard University for 
hospitality while part of this work was performed. 
This work has been supported by the Bundesministerium f\"ur 
Bildung and Forschung (BMBF) grant nos. 05H15WOCAA 
and 05H18WOCA1, and by the Dutch National Organization for Scientific 
Research (NWO).

\begin{appendix}
\section{\boldmath Anomalous dimensions 
of the soft functions}

In this Appendix, we consider the renormalization of the soft functions
that contribute to the renormalization group equations for the 
NLP DY cross section in the $q\overline{q}$ 
channel, and calculate their one-loop anomalous dimensions. 
The general form of the RGE reads 
\begin{equation}
\frac{d}{d\ln\mu}S_{A}\left(\Omega,\omega_{i}\right)=
\sum_{B}\int d\Omega'd\omega_{j}'\,\Gamma_{AB}
\left(\Omega,\omega_{i};\Omega',\omega_{j}'\right)S_{B}
\left(\Omega',\omega'_{j}\right),
\label{eq:admkernel}
\end{equation}
with the anomalous dimension matrix defined in terms of the 
$Z$-factor \eqref{eq:softfnz} as
\begin{equation}
\Gamma_{AB}\left(\Omega,\omega_{i};\Omega',\omega_{j}'\right)=
\sum_{C}\int d\Omega''d\omega''_{k}\,\frac{dZ_{AC}
\left(\Omega,\omega_{i};\Omega'',\omega_{k}''\right)}{d\ln\mu}Z_{CB}^{-1}
\left(\Omega'',\omega''_{k};\Omega',\omega_{j}'\right)\,.
\label{eq:Gamma}
\end{equation}
The inverse Z-factor is defined such that
\bea
&&\sum_{C}\int d\Omega''d\omega''_{k}Z_{AC}
\left(\Omega,\omega_{i};\Omega'',\omega_{k}''\right)Z_{CB}^{-1}
\left(\Omega'',\omega''_{k};\Omega',\omega_{j}'\right)
\nn\\
&&=\delta_{AB}\delta\left(\Omega-\Omega'\right)
\delta\left(\omega-\omega'\right)\,,
\eea
where $\delta\left(\omega-\omega'\right)=\prod_j
\delta\left(\omega_j-\omega_j'\right)$.
For further convenience, we introduce the perturbative expansion of the
renormalization constants
\bea
Z_{AB}\left(\Omega,\omega_{i};\Omega',\omega_{j}'\right)&=&
\delta_{AB}\delta\left(\Omega-\Omega'\right)
\delta\left(\omega-\omega'\right)
+\sum_{n=1}^{\infty}Z_{AB}^{\left(n\right)}
\left(\Omega,\omega_{i};\Omega',\omega_{j}'\right)\,,
\eea
where $Z_{AB}^{\left(n\right)}\propto \alpha_s^n$.

\subsection{Kinematic soft functions}
\label{app:softfnkin}

The basic object from which all kinematic soft functions can be 
derived is the LP position-space soft function generalized to 
arbitrary position $x$,
\begin{equation}
\widetilde{S}_0(x) =
\frac{1}{N_c}\,\mbox{Tr} \,
\langle 0|\mathbf{\bar{T}}(Y^\dagger_+(x) Y_-(x)) 
\,\mathbf{T}(Y^\dagger_-(0) Y_+(0))
|0\rangle\,.
\label{eq:LPgeneralsoftfn}
\end{equation}
The leading-power soft function is recovered for $\vec x =0$, while a 
Taylor expansion in $\vec x$ yields power-suppressed contributions.
Lorentz invariance and invariance under repara\-metrization of the collinear 
basis vectors, $n_{-}\to a n_{-},$ $n_{+}\to a^{-1}n_{+}$, for any $a>0$, 
implies that $\widetilde{S}(x)$ is a function 
\begin{equation}
\widetilde{S}_0(x)=\widetilde{S}_0\left(
\ln(-\np x\nm x \mu^2), \frac{x^2}{n_{+}xn_{-}x}\right)
\end{equation}
of two dimensionless variables. The soft function is an example of a 
closed Wilson loop with two cusps, light-like segments and no 
crossing point, whose anomalous dimension arises only from the cusp 
points (see, e.g., \cite{Brandt:1981kf,Korchemsky:1987wg,
Korchemskaya:1992je,Korchemsky:1993uz}). In particular, they are 
renormalized multiplicatively in position space.

At the one-loop order in dimensional regularization 
with $d=4-2\epsilon$, the bare soft function must
have a simple dependence 
\begin{equation}
\tilde{S}_{0,\rm bare}\left(x\right) = 
1+\frac{\alpha_s}{\pi}\left(-n_{-}xn_{+}x\mu^{2}\right)^{\epsilon}
f\left(\epsilon, \frac{x^2}{n_{+}xn_{-}x}\right)
\end{equation}
on the position variable with some function $f(\epsilon,u)$. The 
explicit evaluation gives
\begin{eqnarray}
\widetilde{S}_{0,\rm bare}(x) &=& 1 + 
\frac{\alpha_s C_F}{\pi}\,
\frac{\Gamma\left(1-\epsilon\right)}{\epsilon^{2}}e^{-\epsilon\gamma_{E}}
\nonumber\\ 
&&\times \left(-\frac{1}{4}n_{-}xn_{+}x\mu^{2}e^{2\gamma_{E}}
\right)^{\epsilon}
\left(\frac{x^2}{n_{-}xn_{+}x}\right)^{1+\epsilon}
{}_{2}F_{1}\left(1,1,1-\epsilon;1-\frac{x^2}{n_{-}xn_{+}x}\right)
\nonumber\\
&=& 1+\frac{\alpha_s C_F}{\pi}\left(
\frac{1}{\epsilon^2} + \frac{L}{\epsilon} + 
\frac{L^2}{2}+\frac{\pi^2}{12} + 
\mbox{Li}_2\left(1-\frac{x^2}{n_{-}xn_{+}x}\right)
+\mathcal{O}(\epsilon)\right),\quad
\label{eq:oneloopkin}
\end{eqnarray}
(see also \cite{Li:2011zp}), 
where we defined
\begin{equation}
L\equiv \ln\left(-\frac{1}{4}n_{-}xn_{+}x\mu^{2}e^{2\gamma_{E}}\right)\,.
\end{equation}
The renormalized position-space soft function satisfies the 
renormalization group equation
\begin{equation}
\frac{d}{d\ln\mu}\,\widetilde{S}_0(x) = 
\Big[2 \Gamma_{\rm cusp} L - 2 \gamma_W \Big]\,
\widetilde{S}_0(x)
\label{eq:rgeS0}
\end{equation}
with $\Gamma_{\rm cusp}(\alpha_s)$ defined in \eqref{eq:adhard}, and 
\begin{equation}
\gamma_W = \mathcal{O}(\alpha_s^2)
\end{equation}
at the one-loop order, as follows from \eqref{eq:oneloopkin}.

$\tilde{S}(x)$ is analytic around the point $x_0=(x^0,0,0,0)$ 
(corresponding to $x^2/(\np x\nm x)=1$). We can therefore obtain the 
RGEs for the kinematic soft functions by 
taking the appropriate derivatives of \eqref{eq:rgeS0} and 
their one-loop expressions from \eqref{eq:oneloopkin}. Let us 
therefore define the following vector of NLP position-space 
soft functions 
\begin{equation}
\vec{S}(x^0) = 
\left(
\widetilde{S}_\perp,
\widetilde{S}_3,
\widetilde{S}_{K2},
\widetilde{S}_{K3},
\widetilde{S}_{x_0}
\right)^T\,,
\end{equation}
where (assuming the transverse plane to be the $x_1$-$x_2$-plane)
\begin{eqnarray}
&& \widetilde{S}_{\perp}(x_0) = 
\frac{ix_{0}}{2}\vec{\partial}^{\,2}_{\perp} 
\widetilde{S}_0(x)_{|\vec{x}=0}\,,
\\
&&\widetilde{S}_3(x_0) = \frac{ix_{0}}{2}\vec{\partial}^{\,2}_3 
\widetilde{S}_0(x)_{|\vec{x}=0}\,,
\\
&&\widetilde{S}_{K2}(x_0) = \frac{3}{4}\,(2i)^2 \,
\partial_{x^0}^{2\,}\!\left[\frac{ix^0}{2} \widetilde{S}(x_0)\right]
\\
&&\widetilde{S}_{K3}(x_0) = 2i\partial_{x^0}\widetilde{S}(x_0)
\\
&&\widetilde{S}_{x_0}(x_0) = \frac{-2i}{x^0-i\varepsilon}\,
\widetilde{S}(x_0)\,.
\end{eqnarray}
$\widetilde{S}_{K2}(x^0)$, $\widetilde{S}_{K3}(x^0)$ and 
$\widetilde{S}_{x_0}(x^0)$ represent the Fourier transforms to 
position space of the kinematic and $x^0$ soft functions 
\eqref{eq:sk2}, \eqref{eq:sk3}, and 
\eqref{eq:x0softfn}, respectively, while 
$\widetilde{S}_{K1}(x_0) = \widetilde{S}_{\perp}(x_0) + 
\widetilde{S}_3(x_0)$. With these definitions, it follows from 
\eqref{eq:rgeS0} that the NLP soft functions above satisfy 
the system of RGEs given by
\begin{equation}
\frac{d}{d\ln\mu}\,\vec{S}(x^0) = 
\Big[2 \Gamma_{\rm cusp} L_0 - 2 \gamma_W \Big]\,
\mathbf{1}\,\vec{S}(x)
+ \Gamma_{\rm cusp}
\left(\begin{array}{ccccc}
0 & 0 &0&0&0 \\
0 & 0 &0&0&+1 \\
0 & 0 &0&-6 &+3 \\
0 & 0 &0&0&-4 \\
0 & 0 &0&0&0 
\end{array}\right)
\vec{S}(x^0)\,,
\label{eq:rgeSNLPkin}
\end{equation}
where now
\begin{equation}
L_0\equiv \ln\left(-\frac{1}{4} (x^0)^2\mu^{2}e^{2\gamma_{E}}\right)\,.
\end{equation}
The mixing of $\widetilde{S}_{3}(x^0)$, $\widetilde{S}_{K2}(x^0)$, 
and $\widetilde{S}_{K3}(x^0)$ into $\widetilde{S}_{x_0}(x^0)$ arises 
from the expansion 
\begin{equation}
L = L_0 - \frac{(x^3)^2}{(x^0)^2} +\mathcal{O}
\left(\frac{(x^3)^4}{(x^0)^4}\right)
\end{equation}
and from $x^0$-derivatives on $L_0$, and is given by the 
cusp anomalous dimension to all orders. The diagonal entries of the 
anomalous dimension matrix are identical for all functions and 
equal to the one of the LP DY soft function. 

We are now in the position to justify the statement made in the 
main text that in the sum of the kinematic corrections to 
$\Delta(z) = \hat{\sigma}(z)/z$ the leading NLP logarithms 
cancel to all orders in perturbation theory. From \eqref{eq:rgeSNLPkin} 
we obtain, by summing,  
\begin{equation}
\frac{d}{d\ln\mu}\,\widetilde{S}_{K1+K2+K3}(x^0) = 
\Big[2 \Gamma_{\rm cusp} L - 2 \gamma_W \Big]\,
\widetilde{S}_{K1+K2+K3}(x^0) - 6\,\Gamma_{\rm cusp} 
\,\widetilde{S}_{K3}(x^0)\,,
\label{eq:skinall}
\end{equation}
that is, the mixing into the $1/x^0$ soft function vanishes. 
Since the kinematic soft functions  
all start at the one-loop order, the evolution can only 
produce terms at most of the order of $\alpha_s (\alpha_s \ln^2(1-z))^n$, 
which correspond to next-to-leading logarithms. 

This conclusion does not hold for $\hat{\sigma}(z)$ defined 
as in \eqref{eq:dsigsq2} itself, for which the relevant kinematic soft 
function is the sum $\widetilde{S}_{K1+K2}(x^0)$. For this 
reason, and for further use in Sec.~\ref{sec:resum}, 
it is instructive to solve the RGE system 
\begin{equation}
\frac{d}{d\ln\mu}
\left(\begin{array}{c}
\widetilde{S}_{\rm NLP}(x^0,\mu) \\
\widetilde{S}_{x_0}(x^0,\mu)
\end{array}\right)
=\left(\begin{array}{cc}
2 \Gamma_{\rm cusp} L_0 - 2 \gamma_W & \Delta \\
0 & 2 \Gamma_{\rm cusp} L_0 - 2 \gamma_W
\end{array}\right)
\left(\begin{array}{c}
\widetilde{S}_{\rm NLP}(x^0,\mu)\\
\widetilde{S}_{x_0}(x^0,\mu)
\end{array}\right).
\label{eq:rgeSNLP2by2}
\end{equation}
For the present case of interest $\widetilde{S}_{\rm NLP}(x^0) = 
\widetilde{S}_{K1+K2}(x^0)$ and the off-diagonal anomalous dimension 
is given by $\Delta=\Gamma_{\rm cusp}$.
Since the RGE is local in position space, for fixed $x^0$ it is 
of the same form as \eqref{eq:hardRGE}, except for the off-diagonal 
term. The general solution can be written as 
\begin{eqnarray}
&& \widetilde{S}_{x_0}(x^0,\mu) = U(\mu,\mu_s) 
\widetilde{S}_{x_0}(x^0,\mu_s)\,\\[0.1cm]
&& \widetilde{S}_{\rm NLP}(x^0,\mu) = U(\mu,\mu_s) 
\left[\widetilde{S}_{\rm NLP}(x^0,\mu_s) + 
a_\Delta(\mu,\mu_s) \widetilde{S}_{x_0}(x^0,\mu_s)\right]
\label{eq:snlpevolved}
\end{eqnarray}
defining
\begin{eqnarray}
U(\mu,\mu_s) = \exp\left[-4 S(\mu_s,\mu) + 2 a_{\gamma_W}(\mu_s,\mu) 
\right] \left(-\frac{1}{4} (x^0)^2 \mu_s^2 e^{2\gamma_E}
\right)^{\!2 a_\Gamma(\mu_s,\mu)}
\end{eqnarray}
with $S(\nu,\mu)$, $a_\Gamma(\nu,\mu)$ as given in 
\eqref{eq:Sdef}, \eqref{eq:adef}. $a_{\gamma_W}$, $a_{\Delta}$ are 
defined analogously to $a_{\gamma}$ with the obvious replacement of 
$\gamma$ by the respective anomalous dimension function.

The $1/x^0$ soft function does not appear directly in the NLP 
DY cross section, but only through mixing. Also, the 
initial condition $\widetilde{S}_{\rm NLP}(x^0,\mu_s)$ 
in \eqref{eq:snlpevolved} is $\mathcal{O}(\alpha_s)$, hence 
to LL accuracy, we may approximate
\begin{equation}
\widetilde{S}_{\rm NLP}^{\rm LL}(x^0,\mu) = 
 U^{\rm LL}(\mu,\mu_s) \,a_\Delta^{\rm LL}(\mu,\mu_s)\,
 \widetilde{S}_{x_0,\rm tree}(x^0,\mu_s)
\label{eq:LLsol}
\end{equation}
with $ \widetilde{S}_{x_0,\rm tree}(x^0,\mu_s) = -2i/(x^0-i\varepsilon)$ and 
\begin{eqnarray}
U^{\rm LL}(\mu,\mu_s) = \mbox{exp}\left[-4 S^{\rm LL}(\mu_s,\mu)
\right]\,,
\qquad 
a_\Delta^{\rm LL}(\mu,\mu_s) = -\frac{2\Delta^{(0)}}{\beta_0}
\ln\frac{\alpha_s(\mu)}{\alpha_s(\mu_s)}\,.
\label{eq:LLfns}
\end{eqnarray}
The function $S^{\rm LL}(\nu,\mu)$ was defined in \eqref{eq:SLL}, 
and $\Delta(\alpha_s) = \Delta^{(0)} \alpha_s/\pi + 
\mathcal{O}(\alpha_s^2)$. 
In this approximation the momentum space solution for 
$\widetilde{S}_{\rm NLP}^{\rm LL}(\Omega,\mu)$ is simply given by 
substituting $ \widetilde{S}_{x_0,\rm tree}(x^0,\mu_s) \to 
\theta(\Omega)$. 

\subsection{\boldmath 
$S_{2\xi}(\Omega,\omega)$ soft function}
\label{app:softqqbar}

Here we consider the renormalization of the soft function 
$S_{2\xi}(\Omega,\omega)$ defined in \eqref{eq:s2xitraced}, which
arises from the $\mathcal{L}^{(2)}_{2\xi}$ Lagrangian insertion. 
As discussed in Sec.~\ref{sec:soft}, this soft function  
mixes into the $1/x^0$ soft function $S_{x_0}(\Omega)$. We therefore
determine the one-loop anomalous dimension matrix $\Gamma_{AB}$ for 
$A,B=x_{0},2\xi$, more precisely 
the terms required for LL resummation, namely the cusp-logarithmic 
terms in the diagonal entries, and the off-diagonal entries. 

The diagonal entry $\Gamma_{x_{0}\, x_{0}}$ has been shown above 
to be identical to the one of the leading power soft function in 
position space. Written as momentum-space kernel corresponding to 
the definition \eqref{eq:admkernel}, and keeping only 
the cusp-logarithm part, we have 
\be
\Gamma_{x_{0}\, x_{0}}(\Omega,\Omega')=
4\frac{\alpha_sC_F}{\pi} 
\ln\frac{\mu}{\mu_s}\delta\left(\Omega-\Omega'\right)\,,
\ee
where $\mu_s$ is a soft scale of order $\Omega$. Its precise value 
is not needed, since it only affects the non-logarithmic term.  
The off-diagonal term $\Gamma_{2\xi\,x_{0}}$ can be inferred from 
\eqref{eq:z2xix0} to be
\be
\label{eq:gamma2xix0}
\Gamma_{2\xi\, x_{0}}\left(\Omega,\omega;\Omega'\right)=
-\frac{\alpha_sC_F}{\pi}
\delta\left(\Omega-\Omega'\right)\delta\left(\omega\right).
\ee
The $1/x^0$ soft function cannot mix into the non-local 
$2\xi$ soft function, hence $\Gamma_{x_{0}\, 2\xi}=0$. 
We use two different methods to extract the diagonal part 
$\Gamma_{2\xi\,2\xi}$. First, we consider the renormalization of the 
soft operator 
\bea
\mathcal{S}_{2\xi}\left(\Omega,\omega\right)&=&
\int\frac{dx^0}{4\pi}\int\frac{d\left(n_{+}z\right)}{4\pi}
e^{i\left(x^0\Omega-n_{+}z\omega\right)/2}\;
\mathbf{\overline{T}}\left[Y_{+}^{\dagger}\left(x_{0}\right)Y_{-}
\left(x_{0}\right)\right] 
\nn\\
&& {} \times \mathbf{T}\left[Y_{-}^{\dagger}\left(0\right)Y_{+}
\left(0\right)\frac{i\partial_{\perp\mu}}{in_{-}\partial}
\mathcal{B}_+^{\mu}\left(z_{-}\right)\right]\label{eq:S2xiOp}
\eea
instead of the soft function 
\begin{equation}
S_{2\xi}\left(\Omega,\omega\right)=\frac{1}{N_c}
\text{Tr}\left\langle 0\right|\mathcal{S}_{2\xi}
\left(\Omega,\omega\right)\left|0\right\rangle,
\label{eq:S2xi}
\end{equation}
which allows us to extract 
$\Gamma_{2\xi\,2\xi}$ from a one-loop calculation of the matrix element 
$\langle g|{\cal S}_{2\xi}|0\rangle$ involving an
external gluon. Second, we compute the soft function \eqref{eq:S2xi}, 
involving the vacuum matrix element $\langle 0|{\cal S}_{2\xi}|0\rangle$, 
at the two-loop level.

\paragraph{First method}

We generalize \eqref{eq:softfnz} to the corresponding operators, 
\begin{equation}\label{eq:Zop}
\left[\mathcal{S}_{A}\left(\Omega,\omega_{i}\right)\right]_{\text{ren}}
=\sum_{B}\int d\Omega'd\omega_{j}'{\cal Z}_{AB}
\left(\Omega,\omega_i;\Omega',\omega_{j}'\right)
\left[\mathcal{S}_{B}\left(\Omega',\omega_{j}'\right)\right]_{\text{bare}}\,.
\end{equation}
The operator $(\mathcal{S}_{2\xi})_{ab}$ carries open colour indices, and 
in general the renormalization factor $({\cal Z}_{2\xi,2\xi})_{ab,cd}$ has 
a matrix structure with respect to these colour indices. The renormalization 
factor of the soft function is obtained by projecting on the colour singlet 
part,
\be
Z_{2\xi\,2\xi}=\frac{1}{N_c}\sum_{a,c}({\cal Z}_{2\xi\,2\xi})_{aa,cc}\,.
\ee
For the leading $1/\epsilon^2$ pole we find that 
$({\cal Z}_{2\xi\,2\xi})_{ab,cd}\equiv\delta_{ac}\delta_{bd}
{\cal Z}_{2\xi\,2\xi}+{\cal O}(\epsilon^{-1})$
is diagonal in colour indices, such that $Z_{2\xi\,2\xi}=
{\cal Z}_{2\xi\,2\xi}+{\cal O}(\epsilon^{-1})$.
Since for the purpose of LL resummation we are interested in the leading 
pole only, we do not discriminate between $Z_{2\xi\,2\xi}$ and 
${\cal Z}_{2\xi\,2\xi}$ in the following.

\begin{figure}
\begin{centering}
\includegraphics[width=0.55\textwidth]{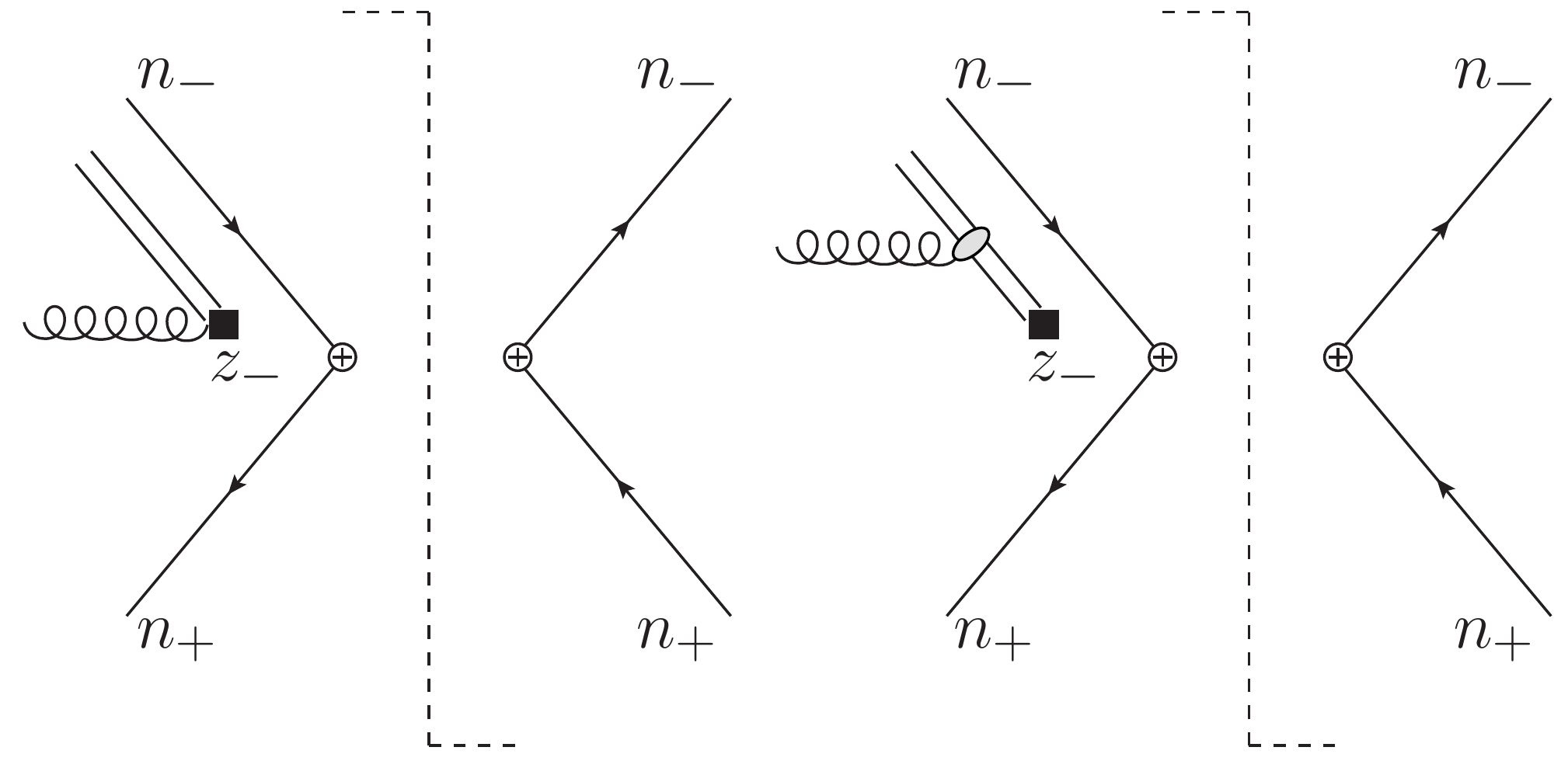}
\par\end{centering}
\caption{\label{fig:Ftree} Tree-level diagrams for the one-gluon matrix
element of the soft operator ${\cal S}_{2\xi}$. The part to the left (right) 
of the cut corresponds to the time-ordered (anti-time-ordered) part of the 
diagram, and lines labeled by $n_\pm$ with in\,(out)-going arrow correspond 
to soft Wilson lines $Y_\mp$ ($Y_\mp^\dag$). The filled square  and the 
two solid lines connected to it stand for the soft covariant
derivative and the Wilson lines contained in 
$\frac{i\partial_{\perp\mu}}{in_{-}\partial}\mathcal{B}_+^{\mu}=
\frac{i\partial_{\perp\mu}}{in_{-}\partial}\,Y_+^\dag[iD_s^{\mu} Y_+]$, 
respectively.
}
\end{figure}

In order to extract $Z_{2\xi\,2\xi}$ we consider a matrix element of 
$\mathcal{S}_{2\xi}$ with a non-vanishing tree-level contribution. We find 
it convenient to consider a single gluon with momentum $p$ and 
colour index $A$ in the external state,
$\langle g_A(p)|\mathcal{S}_{2\xi}|0\rangle$.
The gluon can be emitted from both of the fields inside the time- and 
anti-time ordered part. For a general polarization vector $\epsilon$, 
the tree-level matrix element is (see Fig. \ref{fig:Ftree})
\begin{equation}
\langle g_{A}(p)|\mathcal{S}_{2\xi}(\Omega,\omega)|0\rangle _{
\text{tree}} = g_sT^{A}\left(\frac{p_{\perp}\cdot\epsilon_{\perp}^*}{n_{-}p}
-\frac{p_\perp^2\nm\epsilon^*}{(n_{-}p)^2}\right)
\delta(\Omega)\delta(\omega-n_{-}p).
\end{equation}
The components of the polarization vector are related by
\be
0 = p\cdot\epsilon = 
p_{\perp}\cdot\epsilon_{\perp}+\frac12\nm p\,\np \epsilon 
+\frac12\np p\,\nm \epsilon\,.
\ee

For the computation of the one-loop matrix element, we find it 
convenient to choose a polarization vector satisfying 
$\nm\epsilon=0$, such that the tree-level matrix element is proportional to 
$p_{\perp}\cdot\epsilon_{\perp}^*$ (this implies that only the left diagram
in Fig.~\ref{fig:Ftree} contributes). The one-loop computation 
yields $p_{\perp}\cdot\epsilon_{\perp}$ as well as $\np\epsilon$ terms. 
The latter can be eliminated in terms of the former using
\be
\np\epsilon = -2\,\frac{p_{\perp}\cdot\epsilon_{\perp}}{\nm p}\,.
\ee
The computation is done in dimensional regularization for both 
UV and IR singularities of the on-shell matrix element. As mentioned 
above, it will be sufficient to extract the leading $1/\epsilon^2$ pole.

\begin{figure}[t]
\begin{centering}
\begin{tabular}{ccc}
\includegraphics[width=0.825\textwidth]{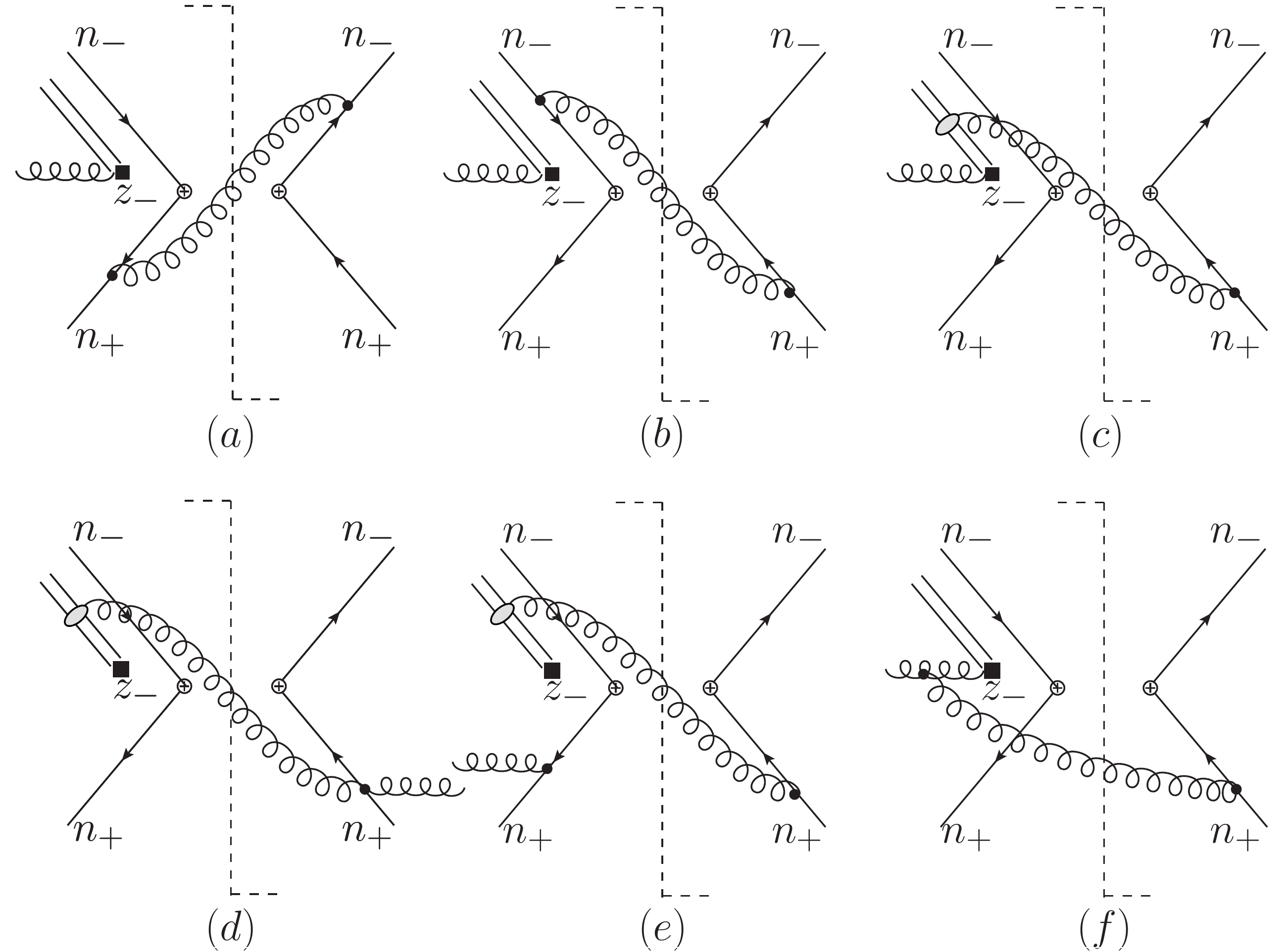} \\
\includegraphics[width=0.825\textwidth]{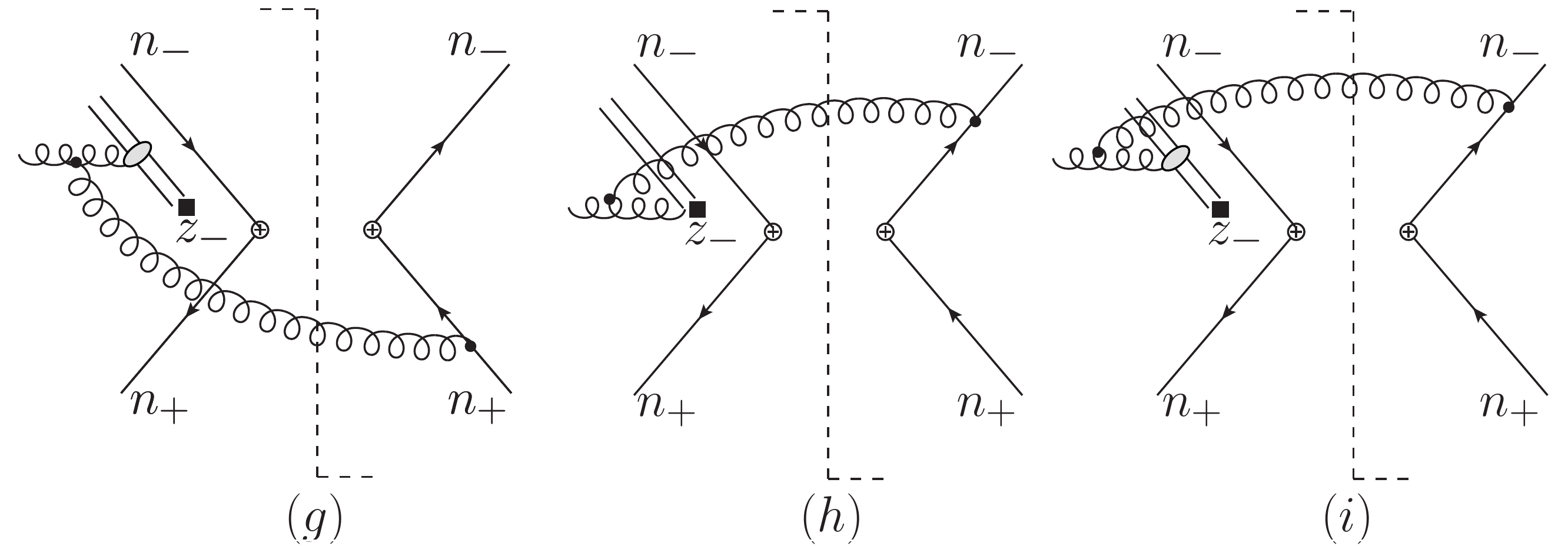} 
\end{tabular}
\par\end{centering}
\caption{\label{fig:Freal} One-loop diagrams with gluon passing the cut
for the one-gluon matrix element of the soft operator ${\cal S}_{2\xi}$.
}
\end{figure}

\begin{figure}[t]
\begin{centering}
\begin{tabular}{ccc}
\includegraphics[width=0.825\textwidth]{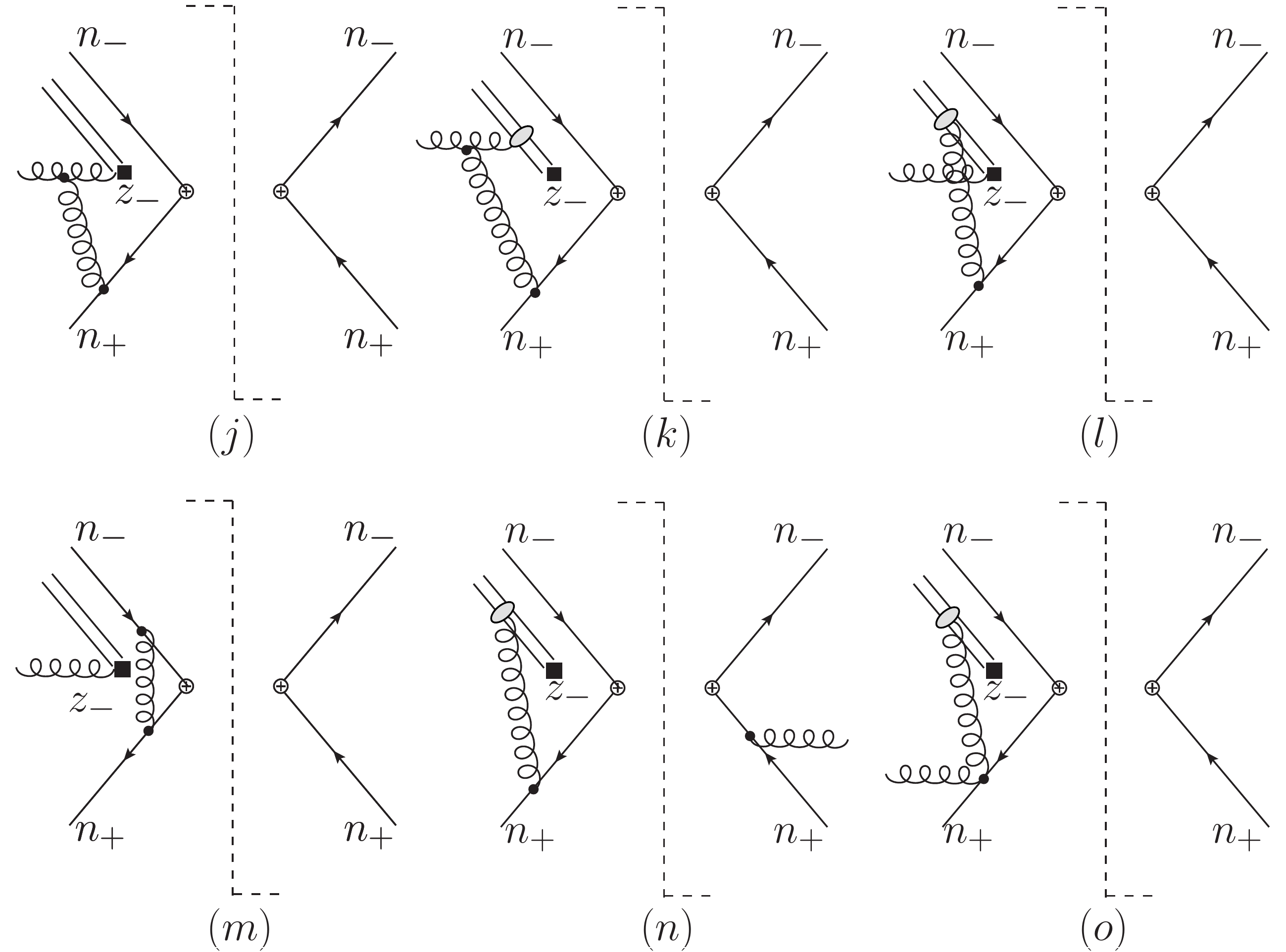} \\
\includegraphics[width=0.825\textwidth]{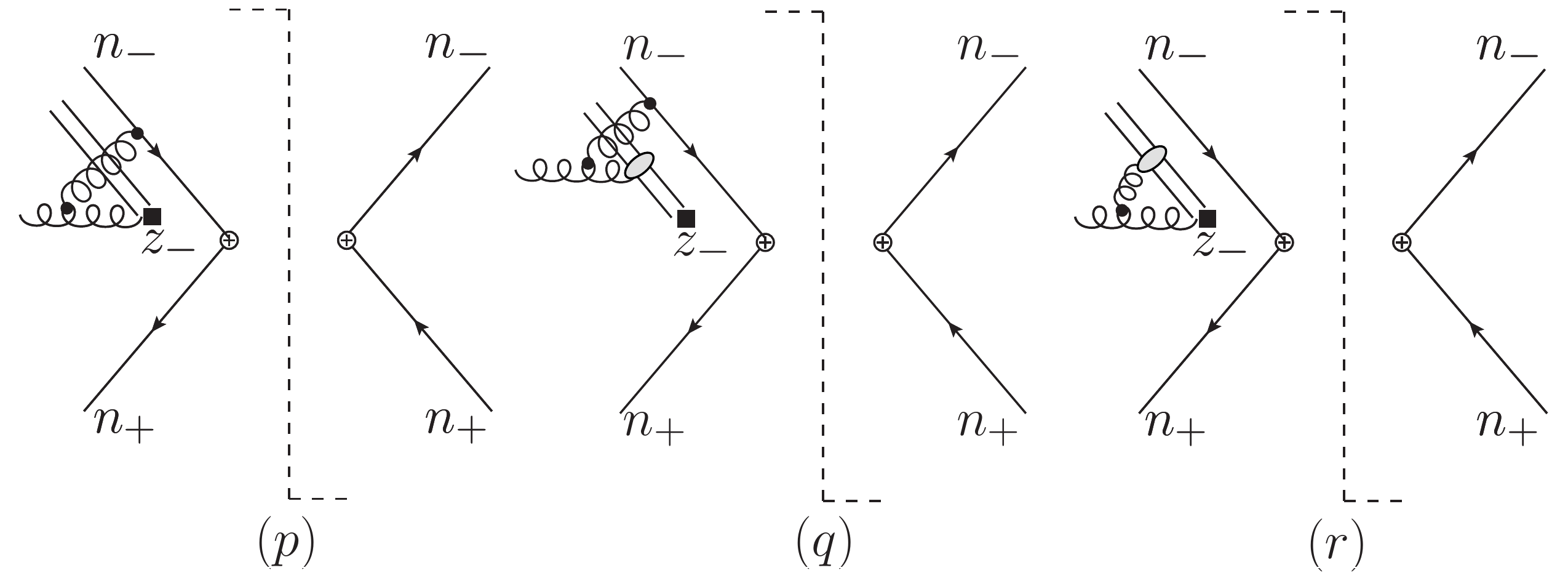} 
\end{tabular}
\par\end{centering}
\caption{\label{fig:Fvirt}One-loop virtual diagrams for the
one-gluon matrix element of the soft operator ${\cal S}_{2\xi}$. 
}
\end{figure}

The relevant diagrams are shown in Figs.~\ref{fig:Freal} and~\ref{fig:Fvirt}.
They can be divided into ``real'' (internal gluon connecting time- and 
anti-time-ordered parts of the operator, see Fig.~\ref{fig:Freal})
and ``virtual'' contributions (internal gluon connecting two fields within 
the time- or anti-time-ordered parts, respectively, see Fig.~\ref{fig:Fvirt}).
In both cases, additional diagrams that vanish
due to our assumption $\nm\epsilon=0$ are not shown. 
In particular, this implies that the external gluon cannot be directly attached
to the Wilson lines contained in ${\cal B}_+^\mu$.
In addition, virtual diagrams with 
an internal gluon line in the anti-time-ordered part are not shown. They 
yield scaleless integrals, analogous to the leading-power soft function.

First consider diagrams $a-c$ in Fig.~\ref{fig:Freal}. In these diagrams, 
the $\perp$ gluon from the $\mathcal{B}^{\mu}_{+}$ insertion is connected 
directly to the external state, while the loop is formed
by the gluons connecting different Wilson lines. We find 
\begin{align}
\langle g_{A}(p)|\mathcal{S}_{2\xi}(\Omega,\omega)|0\rangle_{1\text{-loop}}^
{a)} & 
=\left[\frac{\alpha_{s}}{2\pi}\frac{C_F}{\epsilon^2}
+\mathcal{O}\left(\epsilon^{-1}\right)\right]
\langle g_{A}(p)|\mathcal{S}_{2\xi}(\Omega,\omega)|0\rangle_{\text{tree}}\,,
\nonumber\\
\langle g_{A}(p)|\mathcal{S}_{2\xi}(\Omega,\omega)|0\rangle_{1\text{-loop}}^
{b)} & 
=\left[\frac{\alpha_{s}}{2\pi}\frac{C_{F}}{\epsilon^{2}}
+\mathcal{O}\left(\epsilon^{-1}\right)\right]
\langle g_{A}(p)|\mathcal{S}_{2\xi}(\Omega,\omega)|0\rangle_{\text{tree}}\,,
\nonumber\\
\langle g_{A}(p)|\mathcal{S}_{2\xi}(\Omega,\omega)|0\rangle_{1\text{-loop}}^
{c)} & 
=\left[-\frac{\alpha_{s}}{4\pi}\frac{C_{A}}{\epsilon^{2}}
+\mathcal{O}\left(\epsilon^{-1}\right)\right]
\langle g_{A}(p)|\mathcal{S}_{2\xi}(\Omega,\omega)|0\rangle_{\text{tree}}\,.
\end{align}
Diagrams $d)$ and $e)$ can be shown not to lead to a $1/\epsilon^2$ pole.
Next we compute real diagrams that involve a triple-gluon vertex shown
in Fig.~\ref{fig:Freal} $f-i$. We find that the $1/\epsilon^2$ pole cancels 
among these diagrams. There exist four additional diagrams, that
are obtained from diagrams $f-i$ in the following way: each of these 
diagrams contains two internal gluon lines, one of which is cut. When cutting 
the other gluon line instead, the corresponding diagrams can be shown 
to have no $1/\epsilon^2$ poles.

Finally, we compute the virtual diagrams shown in 
Fig.~\ref{fig:Fvirt}. For diagrams $l)$, $m)$ and $n)$, the internal 
gluon couples only to Wilson lines. We find that they yield scaleless 
integrals, and thus vanish. The same is true for diagrams $p)$, $q)$ and $r)$.
Diagram $o)$ does not yield a $1/\epsilon^2$ pole, similar to $d)$.
Only the first two diagrams $j)$ and $k)$ contribute.
We find 
\begin{equation}
\langle g_{A}(p)|\mathcal{S}_{2\xi}(\Omega,\omega)|0\rangle_{1\text{-loop}}^
{j)+k)}=\left[\frac{\alpha_{s}}{4\pi}\frac{C_{A}}{\epsilon^{2}}
+\mathcal{O}\left(\epsilon^{-1}\right)\right]
\langle g_{A}(p)|\mathcal{S}_{2\xi}(\Omega,\omega)|0\rangle_{\text{tree}}.
\end{equation}
We observe that the $C_A$ term cancels in the sum of real and 
virtual diagrams. The renormalization factor 
is
\begin{equation}
Z_{2\xi\,2\xi}^{\left(1\right)}\left(\Omega,\omega;\Omega',\omega'\right)=
-\frac{\alpha_sC_F}{\pi}\frac{1}{\epsilon^2}
\delta\left(\Omega-\Omega'\right)
\delta\left(\omega-\omega'\right)\,,
\label{eq:Z2xi2xi}
\end{equation}
and therefore the diagonal part of the anomalous dimension matrix is found to 
be
\begin{equation}
\Gamma_{2\xi\,2\xi}\left(\Omega,\omega;\Omega',\omega'\right)=
4\frac{\alpha_s C_F}{\pi}\ln\frac{\mu}{\mu_{s}}
\delta\left(\Omega-\Omega'\right)
\delta\left(\omega-\omega'\right),
\end{equation}
again omitting non-logarithmic,  $\mu$-independent contributions. 

This method of computation does not allow us to determine the exact value 
of $\mu_{s}$ or the non-logarithmic terms associated with the single 
$1/\epsilon$ poles, because we find that the single poles depend on 
the momentum components of the external state. This points to the 
need for a better understanding of the renormalization properties 
of generalized soft functions that arise from time-ordered products 
at subleading power. The issue does not appear to be relevant to 
LL accuracy (the double pole). Nevertheless, to check the above 
calculation, we determine the logarithmic term of $\Gamma_{2\xi\,2\xi}$ 
by another method in the following. 

\paragraph{Second method}

We derive $\Gamma_{2\xi\,2\xi}$ in an alternative way from the two-loop 
calculation of the soft function $S_{2\xi}= \frac{1}{N_c}
\text{Tr}\langle 0|{\cal S}_{2\xi}|0\rangle$. In the following, 
$S_{A}^{(n)}(\Omega,\omega_{i})$ denotes the $\mathcal{O}(\alpha_s^n)$ 
term in the perturbative expansion of the soft function 
$S_{A}(\Omega,\omega_{i})$. 

The renormalization condition \eqref{eq:softfnz} for the two-loop
matrix element, restricted to $A=2\xi, x_{0}$, reads
\begin{equation}
S_{2\xi}^{\left(2\right)}+Z_{2\xi\,x_{0}}^{(1)}S_{x_{0}}^{\left(1\right)}
+Z_{2\xi\,x_{0}}^{\left(2\right)}S_{x_{0}}^{\left(0\right)}
+Z_{2\xi\,2\xi}^{\left(1\right)}S_{2\xi}^{\left(1\right)}
=\text{finite}\,,\label{eq:rencon1}
\end{equation}
where we omit convolutions with respect to $\Omega'$ and $\omega_j'$ for 
brevity. We use this equation to extract the double pole part of 
$Z_{2\xi,2\xi}^{\left(1\right)}$. For that purpose it is sufficient 
to focus on the leading pole $1/\epsilon^3$ term of the equation.

The counterterm $Z_{2\xi\,x_{0}}^{\left(1\right)}$ has already been 
determined in \eqref{eq:z2xix0} by requiring that the one-loop
matrix element \eqref{eq:S2xi0} is finite. We further use that the
renormalization of the $S_{x_0}$ soft function is the same 
as for the leading-power soft function. Thus we have
\begin{equation}
S_{x_0}^{(1)} = - Z_{x_0\,x_0}^{(1)} S_{x_0}^{(0)}
+\mathcal{O}\left(\epsilon^{0}\right)
\end{equation}
and we can rewrite \eqref{eq:rencon1} as
\begin{equation}
S_{2\xi}^{(2)}+\left(-Z_{x_0\,x_0}^{(1)}Z_{2\xi\, x_0}^{(1)}
+ Z_{2\xi\,x_{0}}^{\left(2\right)}\right)S_{x_{0}}^{\left(0\right)}
+Z_{2\xi\,2\xi}^{\left(1\right)}S_{2\xi}^{\left(1\right)}
=\mathcal{O}\left(\frac{1}{\epsilon^{2}}\right),
\end{equation}
where $Z_{2\xi\,x_0}^{(2)}$ and $Z_{2\xi\,2\xi}^{\left(1\right)}$ are 
unknown at this point. Now we use that the relevant entries of $\Gamma_{AA}$ 
have explicit, linear dependence on $\ln\mu$, while the off-diagonal
terms depend on $\mu$ only implicitly through $\alpha_{s}$. Under
these assumptions we can solve \eqref{eq:Gamma} perturbatively 
for $Z_{AB}$ and find a relation that constrains the highest poles of the 
renormalization factor,
\begin{equation}
Z_{AB}^{(2)}=\frac{1}{4}Z_{AB}^{(1)}\left(Z_{AA}^{(1)}
+3 Z_{BB}^{(1)}\right)+\mathcal{O}\left(\frac{1}{\epsilon^{2}}\right),
\;\;A\neq B.
\end{equation}
We could also determine the $1/\epsilon^2$ pole of 
$Z_{AB}^{\left(2\right)}$, but it is not required for LL resummation. 
The above relation implies that the double pole part of the renormalization
constant $Z_{2\xi\,2\xi}^{\left(1\right)}$ can be determined from the 
leading $1/\epsilon^3$ pole of the two-loop vacuum matrix element 
$S_{2\xi}^{\left(2\right)}$ through the equation
\begin{align}
S_{2\xi}^{(2)}-\frac{1}{4} Z_{2\xi\,x_{0}}^{(1)}\left(3 Z_{2\xi\,2\xi}^{
(1)}+Z_{x_0\,x_0}^{(1)}\right) S_{x_{0}}^{\left(0\right)} 
& =\mathcal{O}\left(\frac{1}{\epsilon^{2}}\right).
\end{align}
Due to the implicit convolution, we are not able to reconstruct the 
full dependence on the momentum variables from this equation. Nevertheless, 
for the leading pole, assuming
\be
Z_{2\xi\,2\xi}^{(1)}(\Omega,\omega;\Omega',\omega')\propto
\frac{1}{\epsilon^2}\delta(\Omega-\Omega')\delta(\omega-\omega')
+{\cal O}(1/\epsilon)\,,
\ee
leads to an algebraic equation for the prefactor of the double pole 
that can be solved. We performed the computation of $S_{2\xi}^{(2)}$  
to cross-check the previous method and we obtained a result that agrees 
with \eqref{eq:Z2xi2xi}. 

Combining the results we find the following RGE equation for the 
generalized soft functions at LL accuracy,
\begin{equation}
\frac{d}{d\ln\mu}\left(\begin{array}{c}
S_{2\xi}\left(\Omega,\omega\right)\\
S_{x_{0}}\left(\Omega\right)
\end{array}\right)=\frac{\alpha_{s}}{\pi}\left(\begin{array}{cc}
4C_{F}\ln\displaystyle\frac{\mu}{\mu_s} & -C_{F}\delta(\omega)\\
0 & 4C_{F}\ln\displaystyle\frac{\mu}{\mu_s}
\end{array}\right)\left(\begin{array}{c}
S_{2\xi}\left(\Omega,\omega\right)\\
S_{x_0}\left(\Omega\right)
\end{array}\right)\,,
\label{eq:ADM}
\end{equation}
where $\mu_s={\cal O}(Q(1-z))$.

As discussed in the main text, the ${\cal L}_{2\xi}^{(2)}$ insertion 
appears four times. In addition to the insertion on the incoming 
quark leg, discussed explicitly in the main text, it may also occur along the 
$\np$ direction (``$\bar c$-term''), or within the complex 
conjugated amplitude contributing to the DY cross section, in which case 
the Lagrangian insertion occurs within the anti-time-ordered 
part (``$\bar T$-term''). Finally, there is a contribution from 
${\cal L}_{2\xi}^{(2)}$ along the $\np$ direction
within the conjugated amplitude (``$\bar c\,\bar T$-term'').

These contributions can be treated in a way analogous to the one 
discussed in the main text. The 
$\bar c$-term involves the soft operator
\bea
\mathcal{S}_{2\xi}^{\bar c}\left(\Omega,\omega\right)&=&
\int\frac{dx^0}{4\pi}\int\frac{d\left(n_{-}z\right)}{4\pi}
e^{i\left(x^0\Omega-n_{-}z\omega\right)/2}\,
\mathbf{\bar{T}}\left[Y_{+}^{\dagger}\left(x_{0}\right)Y_{-}
\left(x_{0}\right)\right] 
\nn\\
&& {} \times \mathbf{T}\left[\frac{i\partial_{\perp\mu}}{in_{+}
\partial}
\mathcal{B}_-^{\mu}\left(z_{+}\right)
Y_{-}^{\dagger}\left(0\right)Y_{+}
\left(0\right)\right]\,.\label{eq:S2xiOp_conj}
\eea
The corresponding soft function is 
$S_{2\xi}^{\bar c}\left(\Omega,\omega\right)
=\frac{1}{N_c}{\rm Tr}\langle0|\mathcal{S}_{2\xi}^{\bar c}
\left(\Omega,\omega\right)|0\rangle$. Its value at the one-loop order 
is identical to \eqref{eq:S2xi0} up to a sign, such that 
$\Gamma_{2\xi\,x_0}^{\bar c}=-\Gamma_{2\xi\,x_0}$.
To determine the corresponding diagonal part of the anomalous dimension, 
we proceed as in the first method outlined above. However, we find it 
convenient to use a polarization vector for the external gluon with
$\np\epsilon=0$ and $\nm\epsilon\not=0$. The relevant Feynman diagrams can 
then be obtained from Figs.~\ref{fig:Freal} and~\ref{fig:Fvirt} by 
interchanging $\nm\leftrightarrow \np$ as well as the arrows, and replacing 
$z_-\to z_+$. 
We refer to the labels of the corresponding ``flipped'' diagrams below,
and describe the changes. Due to the different operator ordering, one needs 
to replace $C_F\to C_F-C_A/2$ in $a)$ and $b)$. Diagram $c)$ flips its sign,  
since the virtual gluon is attached to $Y_+^\dag(x_0)$ instead of $Y_-(x_0)$. 
These changes compensate in the sum of $a-c$, which gives the same 
contribution as for $S_{2\xi}$, when expressed in terms of the corresponding 
tree-level matrix element. Apart from that, only the virtual diagrams $j)$ 
and $k)$ contribute to the $1/\epsilon^2$ pole, as before. For these 
contributions, the different colour ordering gives a factor of $(-1)$. In 
addition, since the virtual gluon is attached to $Y_+(0)$ instead of 
$Y_-^\dag(0)$, one obtains another factor of $(-1)$, as well as a different 
sign in the $i\varepsilon$-term arising from the Wilson line. The latter can 
be checked to have no effect on the $1/\epsilon^2$ pole. Therefore, the total 
virtual contribution is also unchanged, and we find 
$\Gamma_{2\xi\,2\xi}^{\bar c}=\Gamma_{2\xi\,2\xi}$. 

This means the
soft functions $(S_{2\xi}^{\bar c},S_{x_0})$ obey an analogous equation as 
$(S_{2\xi},S_{x_0})$, given by \eqref{eq:ADM}, with a sign flip in the 
off-diagonal term.
This sign difference is compensated by a sign flip in the 
corresponding jet function $J_{2\xi}^{(O)}$ appearing in the $\bar c$-term,
such that the contribution to $\Delta$ is the same.

The $\bar T$-term involves the soft operator
\bea
\mathcal{S}_{2\xi}^{\bar T}\left(\Omega,\omega\right)&=&
\int\frac{dx^0}{4\pi}\int\frac{d\left(n_{+}z\right)}{4\pi}
e^{i\left(x^0\Omega+n_{+}z\omega\right)/2}\,
\mathbf{\bar{T}}\left[\frac{i\partial_{\perp\mu}}{in_{-}\partial}
\mathcal{B}_+^{\mu}\left(x_0+z_{-}\right)Y_{+}^{\dagger}\left(x_{0}\right)Y_{-}
\left(x_{0}\right)\right] 
\nn\\
&& {} \times \mathbf{T}\left[
Y_{-}^{\dagger}\left(0\right)Y_{+}
\left(0\right)\right]\,.
\eea
Within our conventions, the sign in  the exponent of $e^{in_{+}z\omega/2}$ 
needs to be flipped compared to $\mathcal{S}_{2\xi}$.
The corresponding soft function $S_{2\xi}^{\bar T}\left(\Omega,\omega\right)
=\frac{1}{N_c}{\rm Tr}\langle0|\mathcal{S}_{2\xi}^{\bar T}\left(\Omega,
\omega\right)|0\rangle$ involves the vacuum matrix element. Since the vacuum 
state is space-time translation invariant, we can add a four-vector $a^\mu$ 
to the argument of all field operators inside this matrix element, without 
changing it. Note that this is true also in presence of the time- and 
anti-time-ordering operators. Choosing $a^\mu= - x_0 = 
(-x^0,0,0,0)$, and then performing
a substitution $x^0\to -x^0$, we find 
\be
S_{2\xi}^{\bar T}\left(\Omega,\omega\right)=S_{2\xi}
\left(\Omega,\omega\right)^*\,,
\ee
where we also used $({\cal B}_+^\mu)^\dag={\cal B}_+^\mu$.
A similar argument yields $S_{x_0}(\Omega)=S_{x_0}(\Omega)^*$ (including 
the $i\varepsilon$-term in \eqref{eq:x0softfn}). 
Therefore, the equations for the soft functions
$(S_{2\xi}^{\bar T},S_{x_0})$ can be obtained by complex conjugation of 
\eqref{eq:ADM}.

By combining the arguments for the $\bar c$- and $\bar T$-terms 
from above, we find that the soft function
arising from the $\bar c\,\bar T$-term satisfies an equation analogous 
to the $\bar c$-term, and the same is true for the jet function. 
In total, all terms contribute equally to $\Delta$.

\section{Derivation of Eq.~(\ref{eq:rge_delta})}
\label{app:derivation}

In this appendix we derive (\ref{eq:rge_delta}) starting from the 
evolution equation \eqref{eq:consistencyDelta},
\begin{equation}
\frac{d}{d\ln\mu}\Delta(z,\mu) = -2
 \int_z^1\frac{dx}{x}
P_{qq}(x)\Delta\left(\frac{z}{x},\mu\right)+{\cal O}(\lambda^4)\,.
\end{equation}
In order to extract the next-to leading power contribution we write
$\Delta(z,\mu)=\Delta_{\rm LP}(z,\mu)+\Delta_{\rm NLP}(z,\mu)+\dots$. 
Eq.\,\eqref{eq:consistencyDelta} then takes the form
\be\label{eq:deltaexpand}
\frac{d}{d\ln\mu}\Delta_{\rm NLP}(z,\mu) = -2
 \int_z^1\frac{dx}{x}
\left(P_{qq}^{\rm LP}(x)\Delta_{\rm LP}
+P_{qq}^{\rm LP}(x)\Delta_{\rm NLP}
+P_{qq}^{\rm NLP}(x)\Delta_{\rm LP}
\right)\big|_{\rm NLP}\,,
\ee
where we indicated that the NLP term should be extracted from the 
right-hand side \emph{after} the integration, and omitted the 
argument of $\Delta_{\rm (N)LP}(z/x,\mu)$ for brevity.
The first term on the right-hand side, expanded to NLP, yields a kinematic 
correction, which we will discuss further below.

After inserting the formal expansions
\bea\label{eq:deltaexp}
\Delta_{\rm LP}(z,\mu) &=& \sum_{n=0}^\infty\alpha_s(\mu)^n
\left\{c_n(\mu)\delta(1-z)
+\sum_{m=0}^{2 n-1}
c_{nm}(\mu)\left[\frac{\ln^m(1-z)}{1-z}\right]_+ \right\},
\nn\\
  \Delta_{\rm NLP}(z,\mu) &=& 
\sum_{n=0}^\infty\alpha_s(\mu)^n\sum_{m=0}^{2 n-1}d_{nm}(\mu) \ln^m(1-z)\,,
\eea
the second term on the right-hand side of 
\eqref{eq:deltaexpand} contains integrals of the form
\bea
\lefteqn{\int_z^1\frac{dx}{x}\left[\frac{1}{1-x}\right]_+\,\ln^m(1-z/x) =}
\nn\\
&=&\ln^{m+1}(1-z)
+\sum_{k=1}^m(-1)^k\zeta(k+1)\frac{m!}{(m-k)!} \ln^{m-k}(1-z)+{\cal O}(1-z)
\nn\\
&=& \left[\ln(1-z)+\sum_{k=1}^m\zeta(k+1)\left(-\frac{d}{d\ln(1-z)}
\right)^{\!k}\right]\,\ln^m(1-z) +{\cal O}(1-z)\,,
\eea
where we used 
\bea\label{eq:plusidentity}
 \int_z^1 dx \,f(x)\left[\frac{1}{1-x}\right]_+ &=& \int_z^1 dx \,
\frac{f(x)-f(1)}{1-x} - \int_0^z dx \,\frac{f(1)}{1-x}\,.
\eea
Note that the upper limit of the summation over $k$ can be extended to 
infinity without changing the result. We can therefore rewrite the result 
in terms of Euler's psi function using
\bea
\sum_{k=1}^\infty(-a)^k\zeta(k+1)&=&-\gamma_E-\psi(1+a)\,.
\eea
This yields
\bea
\lefteqn{  -2\int_z^1\frac{dx}{x} P_{qq}^{\rm LP}(x)\Delta_{\rm NLP}
\left(\frac{z}{x},\mu\right)\big|_{\rm NLP} }\nn\\
  &=& -4\left[\Gamma_{\rm cusp}(\alpha_s)\left(\ln(1-z)
-\gamma_E-\psi\left(1+\frac{d}{d\ln(1-z)}\right)\right)+
\gamma^\phi(\alpha_s)\right]
\Delta_{\rm NLP}(z,\mu)\,.
\nn\\[-0.4cm]
\eea
After substituting $y=z/x$, the last term on the right-hand side of 
\eqref{eq:deltaexpand} gives
\be
-2\int_z^1\frac{dx}{x} P_{qq}^{\rm NLP}(x)\Delta_{\rm LP}
\left(\frac{z}{x},\mu\right)\big|_{\rm NLP} 
  = -2\gamma^{\rm NLP}_{qq}(\alpha_s)\int_z^1\frac{dy}{y}\, 
\Delta_{\rm LP}\left(y,\mu\right)\big|_{\rm NLP} \,.
\ee
In order to extract the strict NLP contribution we should omit any terms of 
NNLP and beyond, and in particular replace $dy/y\to dy$. After this 
replacement, it can be readily checked that the remaining integral  
over $y$ is strictly NLP by inserting \eqref{eq:deltaexp} and using
\be
\int_z^1 dy\,\left[\frac{\ln^m(1-y)}{1-y}\right]_+ = 
\frac{1}{m+1}\ln^{m+1}(1-z)\,.
\ee
Note that \emph{no} terms proportional to powers of $1-z$ appear on the 
right-hand side. A similar argument applies to the
contribution to $\Delta_{\rm LP}$ that is proportional to $\delta(1-z)$.
We therefore obtain
\be
-2\int_z^1\frac{dx}{x} P_{qq}^{\rm NLP}(x)\Delta_{\rm LP}
\left(\frac{z}{x},\mu\right)\big|_{\rm NLP} 
= -2\gamma^{\rm NLP}_{qq}(\alpha_s)\int_z^1 dy \,\Delta_{\rm LP}(y,\mu) \,.
\ee

In order to compute the kinematic correction from the first term on the 
right-hand side of \eqref{eq:deltaexpand}, we use 
\eqref{eq:Pabexpanded} and \eqref{eq:plusidentity}, resulting in 
\bea\label{eq:kincorr}
-2\int_z^1\frac{dx}{x} P_{qq}^{\rm LP}(x)\Delta_{\rm LP}
\left(\frac{z}{x},\mu\right)
&=& -4\left[\Gamma_{\rm cusp}(\alpha_s)\ln(1-z)
+\gamma^{\phi}_{qq}(\alpha_s)\right]\Delta_{\rm LP}\left(z,\mu\right)
\nn\\
&& -\,4\,\Gamma_{\rm cusp}(\alpha_s)I[\Delta_{\rm LP}](z,\mu)\,,
\eea
where
\be
I[\Delta](z,\mu) \equiv \int_z^1 dx \,
\frac{\displaystyle \frac{1}{x}\Delta\left(\frac{z}{x},\mu\right)-
\Delta\left(z,\mu\right)}{1-x}\,.
\ee
The terms in the first line on the right-hand side of \eqref{eq:kincorr} 
do not yield power-suppressed contributions.
In order to treat the integral in the second line, we need to specify
$\Delta_{\rm LP}$. We quote the result from \cite{Becher:2007ty} in the form,
\be\label{eq:DeltaLP}
\Delta_{\rm LP}(z,\mu) = f(Q,\mu_h,\mu_s,\mu)\,
\frac{1}{1-z}\tilde s_{\rm DY}(\partial_\eta,\mu_s)
\left(\frac{Q(1-z)}{\mu_s}\right)^{2\eta}
\left[z^{-\eta}\right]^a\frac{e^{-2\gamma_E\eta}}{\Gamma(2\eta)}\,,
\ee
where $\eta = 2a_\Gamma(\mu_s,\mu)$, 
$f(Q,\mu_h,\mu_s,\mu)=|C_V(-Q^2,\mu_h)^2|U(Q,\mu_h,\mu_s,\mu)$,  
and $\tilde s_{\rm DY}(\partial_\eta,\mu_s)$
as defined in \cite{Becher:2007ty}. The factor $\left[z^{-\eta}\right]^a$ 
was introduced in \cite{Becher:2007ty} (with $a=1$) for convenience.
It does not change the result to LP accuracy, but introduces 
power-suppressed contributions into $\Delta_{\rm LP}$.
Since we collect all NLP contributions in $\Delta_{\rm NLP}$, 
we must omit this factor, that is, set $a=0$ in the following. Using
\be
  I[(1-z)^{-1+2\eta}](z,\mu) = 
  (1-z)^{-1+2\eta}\left[ -\left(\psi(2\eta)+\gamma_E\right)
   +(1-z)+{\cal O}(1-z)^2\right] \,,
\ee
gives
\bea\label{eq:I_Delta_LP}
I[\Delta_{\rm LP}](z,\mu) &=&
f(Q,\mu_h,\mu_s,\mu)\,\frac{1}{1-z}\tilde s_{\rm DY}(\partial_\eta,\mu_s)
\left(\frac{Q(1-z)}{\mu_s}\right)^{2\eta}\frac{e^{-2\gamma_E\eta}}
{\Gamma(2\eta)}
\nn\\
&& \times\, \left[ -\left(\psi(2\eta)+\gamma_E\right)
+(1-z)+{\cal O}(1-z)^2\right]\,.
\eea
In order to unambiguously identify LP and NLP terms, we need to 
specify the power counting order of the various terms for the combined 
expansion in powers of $1-z~\sim \lambda^2$, $\ln(1-z)$ as well as 
$\ln(\mu/Q)$. In particular, $\eta$ formally counts as a single power of a 
scale-dependent logarithm, and therefore all terms involving 
$(1-z)^{2\eta}=\exp[2\eta\ln(1-z)]$ should formally be treated as
contributing to the expansion in logarithms \cite{Becher:2007ty}.\footnote{
This expansion cannot be taken literally, since $\Delta_{\rm LP}$ has to be 
interpreted as a distribution, but it suffices to identify the relevant order 
in power counting.}  This implies that the first term in the square bracket 
in \eqref{eq:I_Delta_LP} should be interpreted as the LP contribution, 
and the second one as the NLP correction. Hence
\be
  I[\Delta_{\rm LP}](z,\mu)|_{\rm NLP}=(1-z)\Delta_{\rm LP}(z,\mu)\,,
\ee
which gives
\bea\label{eq:kincorr_nlp}
 -2\int_z^1\frac{dx}{x} P_{qq}^{\rm LP}(x)\Delta_{\rm LP}
\left(\frac{z}{x},\mu\right)\big|_{\rm NLP}
=  -4\,\Gamma_{\rm cusp}(\alpha_s)(1-z)\Delta_{\rm LP}(z,\mu)\,.
\eea
Note that this result is consistent with the structure of the formal 
expansions \eqref{eq:deltaexp} of $\Delta_{\rm (N)LP}$:
Since the right-hand side contributes to the $\mu$-evolution of 
$\Delta_{\rm NLP}$, it should possess a formal series expansion in terms 
of $\ln^m(1-z)$, $m\geq 1$. Using the formal expansion of $\Delta_{\rm LP}$ 
in the first line of \eqref{eq:deltaexp}, the product  
$(1-z)\Delta_{\rm LP}(z,\mu)$ indeed yields such an expansion,
since, when applied to a regular test function, 
$(1-z)[\ln^m(1-z)/(1-z)]_+=\ln^m(1-z)$, and $(1-z)\delta(1-z)=0$.

One may wonder whether the derivatives contained in 
$\tilde s_{\rm DY}(\partial_\eta,\mu_s)$ in \eqref{eq:I_Delta_LP} implicitly 
contain additional NLP terms. Based on the power counting rules discussed 
above, this is not expected. In order to explicitly check this point, 
we use
\be
\tilde s_{\rm DY}(\partial_\eta,\mu_s)\left(\frac{Q(1-z)}{\mu_s}
\right)^{\!2\eta}
\frac{e^{-2\gamma_E\eta}}{\Gamma(2\eta)} 
  = \left(\frac{Q(1-z)}{\mu_s}\right)^{\!2\eta}\!\tilde s_{\rm DY}
\left(\ln\frac{Q^2(1-z)^2}{\mu_s^2}+\partial_\eta,\mu_s\right)
\frac{e^{-2\gamma_E\eta}}{\Gamma(2\eta)}\,.
\ee
Inserting this into \eqref{eq:DeltaLP}, $\Delta_{\rm LP}$ can formally be 
written as
\be
  \Delta_{\rm LP}(z,\mu) = 
  \sum_{m=0}^{\infty}d_m\,\Delta_m(z,\mu)\,,
\ee
with suitable scale- and $\alpha_s$-dependent, but $z$-independent, 
coefficients $d_m$, and
\be
  \Delta_m(z,\mu)\equiv \frac{\ln^m(1-z)}{(1-z)^{1-2\eta}}\,.
\ee
The integrals relevant for the kinematic correction are
\be
  I_m(z,\mu)\equiv I\left[\Delta_m\right](z,\mu)\,.
\ee
We need to expand the result in powers of $1-z$. We find
\bea
I_m(z,\mu) &=& 
  -\sum_{k=0}^m\left(\psi^{(k)}(2\eta)+\gamma_E\,\delta_{k,0}\right)
\frac{\Gamma(m+1)}{\Gamma(m+1-k)}\Delta_{m-k}(z,\mu)
\nn\\
&& +\,(1-z)\,\Delta_m(z,\mu)+\dots \,,
\eea
where the ellipsis stands for terms of order $(1-z)^2\times\Delta_{k}$, 
$k\geq 0$. The first line of the right-hand side contributes at leading 
power. Using the NLP contribution from the second
line we confirm the previous result \eqref{eq:kincorr_nlp}.

Yet another way to derive \eqref{eq:kincorr_nlp} is \emph{not} to use 
the explicit LP result for $\Delta$, but instead the formal series expansion
\eqref{eq:deltaexp}. We checked explicitly that the result is consistent
with \eqref{eq:kincorr_nlp} up to $m=4$, when inserting the expansion from
the first line of \eqref{eq:deltaexp} on both sides and expanding in
powers of $1-z$ and $\ln(1-z)$, respectively.

Altogether, the evolution equation for the NLP term in the power 
expansion of the partonic cross section reads 
\bea\label{eq:rge_delta2}
\lefteqn{\frac{d}{d\ln\mu}\Delta_{\rm NLP}(z,\mu) } \nn\\
&=&-4\left[\Gamma_{\rm cusp}(\alpha_s) \left(\ln(1-z)
-\gamma_E-\psi\left(1+\frac{d}{d\ln(1-z)}\right)\right)
+\gamma^\phi(\alpha_s)\right]\Delta_{\rm NLP}(z,\mu)\nn\\
&&   + \,K(z,\mu)\,,
\eea
where
\be\label{eq:Kdef2}
  K(z,\mu) \equiv
  -2\gamma^{\rm NLP}_{qq}(\alpha_s)\int_z^1 dy \,\Delta_{\rm LP}
\left(y,\mu\right) 
  -4\Gamma_{\rm cusp}(\alpha_s)(1-z)\Delta_{\rm LP}(z,\mu)\,.
\ee

\bibliography{NLP}

\end{appendix}

\end{document}